\documentclass[fleqn]{2023SCGE}
\usepackage{url}
\usepackage{hyperref}
\usepackage{multirow}
\usepackage[normalem]{ulem}

\newcommand{\apj}{ApJ}
\newcommand{\apjl}{ApJL}
\newcommand{\aj}{AJ}
\newcommand{\apjs}{ApJS}
\newcommand{\mnras}{MNRAS}
\newcommand{\araa}{ARA\&A}
\newcommand{\aap}{A\&A}

\newcommand{\aapr}{A\&A Rv}

\newcommand{\pasp}{PASP}
\newcommand{\pasa}{PASA}
\newcommand{\pasj}{PASJ}

\newcommand{\rmxaa}{RMxAA}

\newcommand{\etal}{et al.}

\newcommand{\kms}{km~s$^{\mathrm{-1}}$}

\newcommand{\MHI}{$M_{\rm H\, \textsc{i}}$}

\newcommand{\msun}{$M_{\mathrm{\odot}}$}

\newcommand{\mstar}{$M_{*}$}

\newcommand{\rmol}{$R_{\rm mol}$\,}
\def\ynk#1 {{\textcolor{cyan}{#1}}\ }
\def\HI{H\,{\textsc{\romannumeral 1}}}
\def\HII{H\,{\textsc{\romannumeral 2}}}
\begin{document}

\ensubject{subject}

%%%%%%%%%%%%%%%%%%%%%%%%%%%%%%%%%%%%%%%%%%%%%%%%%%%%%%%
%%% Authors do not modify the information below
%%% ????????????????
%%% ??????????, ????????????{}, ???????????????????
%Letter to the Editor??Article%??????
\ArticleType{Article}%??Article
\SpecialTopic{SPECIAL TOPIC: Astronomy}%???????
\Year{2024}
\Month{January}
\Vol{xxx}
\No{xx}
\DOI{xxx}
\ArtNo{000000}
\ReceiveDate{\today}
\AcceptDate{\today}
%\OnlineDate{January 1, 2016}
%%%%%%%%%%%%%%%%%%%%%%%%%%%%%%%%%%%%%%%%%%%%%%%%%%%%%%%

%%% title: ????
%%%   \title{title}{title for citation}
\title{The ALMaQUEST Survey XV: The Dependence of the Molecular-to-Atomic Gas Ratios on Resolved Optical Diagnostics}{The ALMaQUEST Survey XV: The Dependence of the Molecular-to-Atomic Gas Ratios on Resolved Optical Diagnostics}
%  in MaNGA Galaxies

% Corresponding author: 
% \author[1]{Niankun Yu}{{niankunyu@bao.ac.cn}}
\author[1, 2]{Niankun Yu$^{\dag}$}{}
% \author[1, 2]{Zheng Zheng}{{niankunyu@bao.ac.cn}}
\author[1, 2]{Zheng Zheng$^{\ddag}$}{}
\author[1, 2, 3, 4]{Chao-Wei Tsai$^*$}{}
\author[1, 2]{Pei Zuo}{}
\author[5]{Sara L. Ellison}{}
% \author[6]{Lihwai Lin}{}
\author[6,7]{\\ David V. Stark}{}
\author[1, 2, 8, 9]{Di Li}{}
\author[4, 1, 2]{Jingwen Wu}{}
\author[6]{Karen L. Masters}{}
\author[10]{Ting Xiao}{}
\author[1, 2, 4]{Yinghui Zheng}{}
\author[1, 2]{\\Zongnan Li}{}
\author[1, 2]{Kai Zhang}{}
\author[1, 2]{Hongying Chen}{}
\author[1, 2]{Shu Liu}{}
\author[1, 2]{Sihan Jiao}{}
\author[4, 1, 2]{Fanyi Meng}{}
\footnote{$^{\dag}$ niankunyu@bao.ac.cn}
\footnote{$^{\ddag}$ zz@bao.ac.cn}
\footnote{$^*$ cwtsai@nao.cas.cn}

\AuthorMark{Yu N.}%\authorcr????????

%%% Authors for citation. ????????§Ö????????
%%% ??????????????, ??????????author???
\AuthorCitation{Yu N., Zheng Z., Tsai C.-W., and et al.}

%%% Address. ???
%%%   \address[number]{Address, City {\rm Postcode}, Country}
\address[{\rm1}]{National Astronomical Observatories, Chinese Academy of Sciences, Beijing, 100101,  P.R. China}
\address[{\rm2}]{Key Laboratory of Radio Astronomy and Technology, Chinese Academy of Sciences, Beijing, 100101, P.R. China}
\address[{\rm3}]{Institute for Frontiers in Astronomy and Astrophysics, Beijing Normal University, Beijing, 102206, P.R. China}
\address[{\rm4}]{School of Astronomy and Space Science, University of Chinese Academy of Sciences, Beijing, 100049, P.R. China}
\address[{\rm5}]{Department of Physics \& Astronomy, University of Victoria, Victoria, British Columbia, V8P 1A1, Canada}
% \address[{\rm6}]{Institute of Astronomy \& Astrophysics, Academia Sinica, Taipei 10617, Taiwan}
\address[{\rm6}]{Departments of Physics and Astronomy, Haverford College, Haverford, PA 19041, USA}
\address[{\rm7}]{Space Telescope Science Institute, San Martin Dr., Baltimore, MD 21218, USA}
\address[{\rm8}]{Zhejiang Laboratory, Hangzhou, Zhejiang Province, 311121, P.R. China}
\address[{\rm9}]{New Cornerstone Science Laboratory, Shenzhen, Guangdong Province, 518054, China}
\address[{\rm10}]{ School of Physics, Zhejiang University, Hangzhou, Zhejiang Province, 310058, P.R. China}

%\contributions{}%????????

%%% Abstract. ??
\abstract{The atomic-to-molecular gas conversion is a critical step in the baryon cycle of galaxies, which sets the initial conditions for subsequent star formation and influences the multi-phase interstellar medium. We compiled a sample of 94 nearby galaxies with observations of multi-phase gas contents by utilizing public \HI, CO, and optical IFU data from the MaNGA survey together with new FAST \HI\ observations. In agreement with previous results, our sample shows that the global molecular-to-atomic gas ratio (\rmol\ $\equiv$ log $M_{\rm H_2}$/\MHI) is correlated with the global stellar mass surface density $\mu_*$ with a Kendall's $\tau$ coefficient of 0.25 and $p < 10^{-3}$, less tightly but still correlated with stellar mass and NUV$-$ r color, and not related to the specific star formation rate (sSFR).
The cold gas distribution and kinematics inferred from the \HI\ and CO global profile asymmetry and shape do not significantly rely on $R_{\rm mol}$. 
Thanks to the availability of kpc-scale observations of MaNGA, we decompose galaxies into \HII, composite, and AGN-dominated regions by using the BPT diagrams. 
% The fraction of \HII\ regions within 1.5 effective radius drops with increasing $R_{\rm mol}$, mainly due to the increasing of metallicity and change in ionization states. 
With increasing $R_{\rm mol}$, the fraction of \HII\ regions within 1.5 effective radius decreases slightly; the density distribution in the spatially resolved BPT diagram also changes significantly, suggesting changes in metallicity and ionization states.
Galaxies with high $R_{\rm mol}$ tend to have high oxygen abundance, both at one effective radius with a Kendall's $\tau$ coefficient of 0.37 ($p < 10^{-3}$) and their central regions. Among all parameters investigated here, the oxygen abundance at one effective radius has the strongest relation with global $R_{\rm mol}$. The dependence of gas conversion on gas distribution and galaxy ionization states is weak. In contrast, the observed positive relation between oxygen abundance ($\mu_*$) and $R_{\rm mol}$ indicates that the gas conversion is efficient in regions of high metallicity (density).}%ÕªÒª

%%% Keywords. ?????
%%%%%%%%% https://s3-us-west-2.amazonaws.com/clarivate-scholarone-prod-us-west-2-s1m-public/wwwRoot/prod3/societyimages/scpma/pacs.pdf
%%%%%%%%% 98.52.–b Normal galaxies; extragalactic objects and systems by type
%%%%%%%%% 98.58.–w Interstellar medium ISM and nebulae in external galaxies
%%%%%%%%% 95.30.Ky Atomic and molecular data, spectra, and spectral parameters ~opacities, rotation constants, line identification, oscillator strengths, gf values, transition probabilities, etc.!
%%%%%%%%% Hydrogen —21-cm lines —external galaxies, 98.58.Ge
%%%%%%%%% 98.58.Bz Atomic, molecular, chemical, and grain processes
%%%%%%%%% 98.58.Hf, H II regions; emission and reflection nebulae
%%%%%%%%% 98.62.Ai Origin, formation, evolution, age, and star formation
%%%%%%%%% 32.30.–r Atomic spectra
%%%%%%%%% 32.70.–n Intensities and shapes of atomic spectral line
%%%%%%%%% 
%%%%%%%%% 
\maketitle

%\tableofcontents%?????

%%%%%%%%%%%%%%%%%%%%%%%%%%%%%%%%%%%%%%%%%%%%%%%%%%%%%%%
%%% The main text. ???????
%???????????????????\cref{fig1}
%\twocolumn\onecolumn
%%%%%%%%%%%%%%%%%%%%%%%%%%%%%%%%%%%%%%%%%%%%%%%%%%%%%%%
\begin{multicols}{2}

\section{Introduction}
\label{sec:intro}
Gas is the raw material that forms stars in galaxies, which plays a crucial role in galaxy formation and evolution (e.g., \cite{Kennicutt1998ApJ...498..541K, Leroy2008AJ....136.2782L, Kennicutt2012ARA&A..50..531K, Lilly2013ApJ...772..119L, Saintonge2022ARA&A..60..319S}).
The cold gas predominantly consists of neutral atomic (\HI) and molecular hydrogen (H$_2$). Within the baryon cycle, atomic gas collapses to form dense molecular gas, which directly fuels star formation in galaxies. The tight relations between star formation and molecular gas, in terms of surface density (Schmidt-Kennicutt law, \cite{Schmidt1959ApJ...129..243S, Kennicutt1998ApJ...498..541K, Leroy2008AJ....136.2782L, Lin2019ApJ...884L..33L, Ellison2021MNRAS.501.4777E, Pessa2021A&A...650A.134P}) and global mass (e.g, \cite{Gao2004ApJ...606..271G, Saintonge2017ApJS..233...22S}), have been extensively studied, but insufficient research has been conducted to adequately investigate the connection between the atomic-to-molecular gas conversion process (\HI-H$_2$) and factors such as metallicity or ionization states.

The \HI-H$_2$ conversion primarily depends on the mid-plane pressure \cite{Elmegreen1993ApJ...411..170E, Blitz2004ApJ...612L..29B, Blitz2006ApJ...650..933B}, radiation field \cite{Elmegreen1993ApJ...411..170E}, and metallicity \cite{Krumholz2009ApJ...693..216K}. The dominant factor is the mid-plane pressure, which is proportional to the surface density (both gas and star) after assuming a thin disk with balanced gas and stars \cite{Blitz2004ApJ...612L..29B}. The radiation field influences the molecular gas dissociation, and metals/dust act as the catalyst in the \HI-H$_2$ conversion. Leroy \etal\, \cite{Leroy2008AJ....136.2782L} found that the surface density ratio of H$_2$ and \HI\ for spiral and dwarf galaxies increases with the increasing of local stellar mass surface density and mid-plane pressure, while it decreases with the increasing of galactic radius. If the \HI\ surface density exceeds $\sim$ 9 \msun\ pc$^{-2}$ at solar metallicity, molecular hydrogen can stably exist against self-shielding, as supported by both theoretical \cite{Krumholz2009ApJ...693..216K, Krumholz2009ApJ...699..850K} and observational \cite{Bigiel2008AJ....136.2846B} evidence. Moreover, both semi-analytic models \cite{Krumholz2009ApJ...693..216K, Popping2014MNRAS.442.2398P} and hydro-dynamical simulations \cite{Lagos2011MNRAS.418.1649L} show consistent results.

The global values of molecular-to-atomic gas ratio ($R_{\rm mol}\equiv$ log $M_{\rm H_2}$/\MHI) describe the efficiency of gas conversion and constrain the physical properties of the interstellar medium (ISM, \cite{Boselli2014A&A...564A..66B}). \rmol\ shows a weakly increasing trend with the global stellar mass \mstar, stellar mass surface density $\mu_*$, and NUV$-$r color for galaxies with log\,(\mstar/\msun) $\geq$ 10.0 \cite{Saintonge2011MNRAS.415...32S}. These relations become stronger while extending to the low mass end and with a larger sample \cite{Bothwell2014MNRAS.445.2599B, Boselli2014A&A...564A..66B, Catinella2018MNRAS.476..875C}. Stark \etal\, \cite{Stark2013ApJ...769...82S} found a positive relation between \rmol\ and enhancements of central star formation in spirals, but the enhancement is not applicable for massive early-type galaxies. 
If $\mu_* \gtrsim 10^{8.7}$ \msun\ pc$^{-2}$, the cold gas reservoir is depleted or dispelled; otherwise, \rmol\ is positively correlated with $\mu_*$ \cite{Saintonge2022ARA&A..60..319S}. 
Additionally, the spatial distribution of \HI\ and H$_2$ may play a crucial role in their conversion. 

The optical diagnostics inferring galaxy ionization states and metallicity may play an important role in understanding the physics of gas conversion. Ionization states of gas in a galaxy can be inferred through optical diagnostics such as the BPT diagrams \cite{BPT1981PASP...93....5B}. These diagnostic tools characterize the ionization states by using strong optical emission lines, such as H $\alpha\ \lambda$ 6564,  H $\beta\ \lambda$ 4862, [S {\scriptsize{II}}] $\lambda\lambda$ 6716, 6731, [N {\scriptsize{II}}] $\lambda\lambda$ 6548, 6583, and [O {\scriptsize{III}}] $\lambda\lambda$ 4959, 5007. For instance, the line emission from galaxies or regions can be categorized into \HII, composite, and AGN-dominant emissions based on their positions in the [O {\scriptsize{III}}]/H$\beta$ versus [N {\scriptsize{II}}]/H$\alpha$ ([N {\scriptsize{II}}] BPT) plane following Kewley \etal\, \cite{Kewley2001ApJ...556..121K} and Kauffmann \etal\, \cite{Kauffmann2003MNRAS.346.1055K} or using the P1P2 BPT diagram following Ji \& Yan \cite{Ji2020MNRAS.499.5749J}. Moreover, the [S {\scriptsize{II}}]-based BPT ([O {\scriptsize{III}}]/H$\beta$ versus [S {\scriptsize{II}}]/H$\alpha$) classifies galaxies or regions into \HII, Low-Ionization Nuclear Emission Line Regions (LINERs), and Seyfert \cite{Kewley2006MNRAS.372..961K}. In some special cases, shocks can show up even in the star-forming regions of the BPT diagrams \cite{Allen2008ApJS..178...20A}. Typically, \HII\ emission-dominated galaxies or regions primarily exhibit emission from massive young stars, while AGN-dominant regions are driven by AGN photoionization or shock, and composite components are results of a mixture of star formation, shock excitation, and/or AGN activity \cite{Kewley2019ARA&A..57..511K}.

Theoretical models take metallicity as a controlling parameter in H$_2$ formation \cite{Elmegreen1993ApJ...411..170E, Krumholz2009ApJ...693..216K, Popping2014MNRAS.442.2398P}, which is supported by observations in nearby galaxies. \rmol\ and gas metallicity are tightly correlated 
%has a Pearson correlation coefficient of 0.5 
\cite{Boselli2014A&A...564A..66B}, because metallicity serves as an indicator of dust content, which facilitates the \HI-H$_2$ conversion as a catalyst \cite{Krumholz2009ApJ...693..216K}. At a given stellar mass, a high star formation rate (SFR) is associated with low metallicity \cite{Ellison2008ApJ...672L.107E, Curti2020MNRAS.491..944C} but a high both atomic and molecular gas fraction \cite{Saintonge2016MNRAS.462.1749S}. Thus the complex effects of gas content and metallicity on the atomic-to-molecular gas conversion are non-trivial.
% However, a high SFR is associated with low metallicity at a given stellar mass \cite{Ellison2008ApJ...672L.107E, Curti2020MNRAS.491..944C}, thus we would expect that efficient atomic-to-molecular gas conversion and abundant gas reservoirs correspond to a lower metallicity. 
The influence of the metallicity distribution on the \HI-H$_2$ transition remains unknown. So more studies are critical to reveal the connection between gas conversion and metallicity.

Detailed studies on this field require high-resolution optical integral field unit (IFU) data and cold gas data with a large sample size. The Mapping Nearby Galaxies at APO (MaNGA, \cite{Bundy2015ApJ...798....7B_MaNGA, Abdurrouf2022ApJS..259...35A}) survey and nearby cold gas surveys provide spatially resolved metallicity and gas content information for nearby galaxies. The MaNGA~project utilizes IFU spectroscopy to map the detailed composition and kinematic structures of 10,010 nearby galaxies. Its \HI\ follow-up (\HI-MaNGA: \cite{Masters2019MNRAS.488.3396M, Stark2021AAS...23752707S}) collects 3669 unique \HI\ observations from GBT and Arecibo. Moreover, the ALMA MaNGA~Quenching \& Star-Formation Survey (ALMaQUEST) utilizes high-resolution spatially resolved optical spectroscopy from MaNGA~and ${^{12}}{}\rm{CO}$ (J = 1-0) follow-up observations with ALMA in the C43-2 configuration to study stellar and gas properties on the same physical scales \cite{Lin2019ApJ...884L..33L, Lin2020ApJ...903..145L}. The IFU and CO observations of MaNGA~and ALMA have a beam size of 2.5$^{\prime \prime}$  (1$^{\prime \prime}\approx$ 1 kpc) with 0.02$\lesssim z \lesssim$0.05 and a field of view of 50$^{\prime \prime}$. For galaxies with MaNGA IFU, CO, and \HI\ data, we could investigate the relation between gas conversion and optical diagnostics. However, the ALMaQUEST survey lacks complete and homogeneous \HI\ observations.

In this paper, we present new observations of \HI\ global spectra for the 37 galaxies in the ALMaQUEST survey using the Five-hundred-meter Aperture Spherical Telescope (FAST, \cite{Nan2011IJMPD..20..989N}, \cite{Li2018IMMag..19..112L}). To build a large sample for statistical analysis, we match the MaNGA~survey with ALMaQUEST, the CO follow-up of the GALEX Arecibo SDSS survey (xCOLD GASS: \cite{Saintonge2011MNRAS.415...32S, Saintonge2017ApJS..233...22S}), the APEX Low-redshift Legacy Survey for MOlecular Gas (ALLSMOG: \cite{Bothwell2014MNRAS.445.2599B, Cicone2017A&A...604A..53C}), and ``JCMT dust and gas In Nearby Galaxies Legacy Exploration'' (JINGLE:
\cite{Saintonge2018MNRAS.481.3497S, DeLooze2020MNRAS.496.3668D}) surveys to investigate the dependence of $R_{\rm mol}$ on gas distribution and metallicity. 
 In Section~\ref{sec:sample}, we introduce the sample properties and the cross-match results. The \HI\ data processing and spectral measurements are presented in Section~\ref{sec:data}. We investigate the dependence of molecular-to-atomic gas ratio in Section~\ref{sec:HIH2}. The main results are summarized in Section~\ref{sec:sum}. This work adopts the following parameter for a $\Lambda$CDM cosmology: 
$\Omega_m$= 0.315, $\Omega_{\Lambda}$= 0.685, and $H_0$ = 67.4 \kms\ Mpc$^{-1}$ \cite{PlanckCollaboration2020AA...641A...6P}. 
We adopt a Chabrier initial mass function (IMF, \cite{Chabrier2003PASP..115..763C}) to uniformly derive SFRs and stellar masses.

\section{The Sample} 
\label{sec:sample}
We compile a sample of galaxies with atomic and molecular gas measurements from \HI\ and CO surveys to construct a comprehensive sample for investigating the atomic-to-molecular gas conversion (\HI-H$_2$). The molecular gas in nearby galaxies has been extensively studied through various large observation programs such as the ALMaQUEST survey \cite{Lin2019ApJ...884L..33L, Lin2020ApJ...903..145L}, xCOLD GASS \cite{Saintonge2011MNRAS.415...32S, Saintonge2017ApJS..233...22S}, ALLSMOG \cite{Bothwell2014MNRAS.445.2599B, Cicone2017A&A...604A..53C}, and the JINGLE survey \cite{Saintonge2018MNRAS.481.3497S, DeLooze2020MNRAS.496.3668D}. The xCOLD GASS sample covers a stellar mass range of $10^{9.0}\ -\ 10^{11.5}$\,\msun\ with a redshift of 0.01$-$0.05 and includes star-forming, green valley, and quenched galaxies. The \HI\ surveys primarily consist of xGASS \cite{Catinella2010MNRAS.403..683C, Catinella2018MNRAS.476..875C} and \HI-MaNGA \cite{Masters2019MNRAS.488.3396M, Stark2021AAS...23752707S}, which combines new GBT observations with existing data from the Arecibo Legacy Fast ALFA (ALFALFA) survey \cite{Haynes2011AJ....142..170H, Haynes2018ApJ...861...49H}. We collect the \HI\ and CO data of our sample galaxies from the datasets listed above.

\subsection{The ALMaQUEST and Supplementary Samples}
\label{subsec:almaquest}

We obtained \sout{the} FAST \HI\ spectral line observations for 37 galaxies, which are selected from 51 galaxies from the ALMaQUEST survey \cite{Lin2020ApJ...903..145L} and supplemented with 2 galaxies (MaNGA~8439-6102 and MaNGA~8250-6104) from Gao et al. \cite{Gao2019ApJ...887..172G}.
These 51 galaxies include 46 published systems in the original ALMaQUEST Survey and 5 additional galaxies from the extended ALMaQUEST merger sample \cite{Thorp2022MNRAS.516.1462T}, the latter of which are either not covered by previous \HI\ observations or have a detection of signal-to-noise ratio (S/N) less than 3. The two galaxies included in Gao et al. \cite{Gao2019ApJ...887..172G} exhibit a gas-rich nature in their molecular phase (CO 2-1) and are actively star-forming. Their stellar masses and redshift fall within similar ranges (10.0$<$ log $M_{\star}$/\msun$<$11.5,  $z\sim$0.04) compared to the ALMaQUEST sample. 
We also include these two systems in our analysis.

These 37 galaxies were not included by \HI-MaNGA data release 1 \cite{Masters2019MNRAS.488.3396M} but fall within the declination range of $-$ 14$^{\circ}$ to +66$^{\circ}$.
This selected sample (referred to as the ALMaQUEST sample in this paper) consists of galaxies with a stellar mass range of $10.0\lesssim$ log (\mstar/\msun) $\lesssim 11.5$ and a redshift range of 0.02 $\lesssim z \lesssim$ 0.13. 
This sample primarily consists of galaxies in the green valley, star-forming main sequence, and star-burst categories \cite{Lin2019ApJ...884L..33L, Lin2020ApJ...903..145L, Lin2022ApJ...926..175L, Ellison2020MNRAS.492.6027E}. We adopt the values of
stellar masses and SFRs by using the Chabrier IMF instead of Salpeter IMF used in 
previous work through utilizing the conversion of Madau \& Dichinson \cite{Madau2014ARA&A..52..415M}. 

To investigate the dependence of \HI-H$_2$ conversion, we also compile a set of galaxies with similar properties to the ALMaQUEST sample. In addition to the ALMaQUEST survey, galaxies with both \HI\ and CO observation are mainly from the xCOLD GASS, ALLSMOG, and JINGLE surveys. Sample galaxies selected from these surveys are with relatively uniform sample selection criteria, accordant global \HI\ and CO profiles, and a large sample size that covers a stellar mass range of 9.0 $\lesssim$ log (\mstar/\msun) $\leq 11.5$. Among these surveys, the xCOLD GASS survey provides both accessible \HI\ and CO spectra. 

We adopt a constant CO(1-0) to H$_2$ conversion factor: $\alpha_{\rm CO}=$ 4.35 \msun\ (K km s$^{-1}$ pc$^2$)$^{-1}$ following Bolatto \etal\, \cite{Bolatto2013ARA&A..51..207B} and references therein. 
Our sample has an oxygen abundance 12+log\,(O/H)\,$\gtrsim$\,8.4 (see details in Section~\ref{subsec:z}), thus the value of $\alpha_{\rm CO}$ does not depend on galaxy metallicity \cite{Sandstrom2013ApJ...777....5S, Bolatto2013ARA&A..51..207B}. For galaxies in the ALLSMOG and JINGLE surveys that only have CO(2-1) observations, we convert the intrinsic brightness luminosity $L_{\rm CO(2-1)}$ to $L_{\rm CO(1-0)}$ by assuming a ratio of $r_{21}$ = $L_{\rm CO(2-1)}/L_{\rm CO(1-0)}$= 0.8 as suggested by Cicone \etal\ \cite{Cicone2017A&A...604A..53C} for nearby normal star-forming galaxies, and the recent spatially resolved studies from Leroy \etal\, \cite{Leroy2022ApJ...927..149L} also give similar but slightly lower values. 
We investigate the dependence of cold gas conversion on galaxy ionization states (Section~\ref{subsec:bpt}) and metallicity (Section~\ref{subsec:z}).

\clearpage
\begin{table*}[!htbp]
\centering
\footnotesize
    \begin{threeparttable}
        \caption{Log of FAST \HI\ Observations for 37 Galaxies}
        \label{tab:observation}
        \centering
        \begin{tabular}{lrrcrcclc}
\\
\hline%
\hline%
Plate-IFU & R.A. & Decl. & $z$ & $D_L$ & \HI-MaNGA & Sample & $T_{\rm int}$ & \HI\ Flag \\
 & (J2000) & (J2000) &  & (Mpc) & & & (seconds) & \\
(1) & (2) & (3) & (4) & (5) & (6) & (7) & (8) & (9) \\
\hline
7815-12705 & 21:15:57.71 & +09:32:35.1 & 0.02955 & 134 & 0 & 0 & 120$\times$3+120$\times$2 & 1 \\
7977-3704 & 22:11:11.70 & +11:48:02.6 & 0.02724 & 123 & 2 & 0 & 120$\times$5+120$\times$3 & 1 \\
7977-9101 & 22:04:29.50 & +12:26:33.5 & 0.02656 & 120 & 2 & 0 & 120$\times$1+120$\times$3 & 1 \\
7977-12705 & 22:11:34.28 & +11:47:45.3 & 0.02724 & 123 & 1 & 0 & 120$\times$1+120$\times$1 & 1 \\
8077-6104 & 02:48:07.87 & $-$00:45:08.3 & 0.04601 & 211 & 1 & 0 & 180$\times$9+180$\times$4 & 1 \\
8077-9101 & 02:46:34.35 & $-$00:50:36.7 & 0.04323 & 198 & 2 & 0 & 180$\times$8+120$\times$4 & 1 \\
8081-3704 & 03:19:17.15 & $-$00:58:10.7 & 0.05400 & 249 & 0 & 0 & 180$\times$8+180$\times$4 & 1 \\
8081-6102 & 03:19:45.63 & $-$00:04:37.9 & 0.03719 & 169 & 2 & 0 & 120$\times$3 & 3 \\
8081-9101 & 03:11:05.32 & $-$00:32:47.5 & 0.02846 & 129 & 2 & 0 & 120$\times$4+120$\times$4 & 1 \\
8081-12703 & 03:21:33.93 & $-$00:10:42.1 & 0.02558 & 116 & 2 & 0 & 120$\times$7+120$\times$4 & 1 \\
8082-12704 & 03:19:47.89 & $-$00:13:16.1 & 0.13214 & 642 & 0 & 0 & 180$\times$8 & 3 \\
8083-12702 & 03:20:58.90 & $-$00:22:03.7 & 0.02104 & 95 & 0 & 0 & 120$\times$1+120$\times$2 & 1 \\
8084-3702 & 03:22:32.79 & $-$00:00:04.4 & 0.02206 & 99 & 2 & 0 & 180$\times$9+180$\times$3 & 1 \\
8084-6103 & 03:22:58.00 & +00:03:14.9 & 0.03593 & 164 & 2 & 0 & 180$\times$8+180$\times$10 & 2 \\
8155-6101 & 03:35:15.39 & $-$01:13:43.0 & 0.03740 & 170 & 1 & 0 & 120$\times$2+120$\times$3 & 1 \\
8155-6102 & 03:30:29.13 & +00:45:07.4 & 0.03081 & 140 & 0 & 0 & 120$\times$5 & 1 \\
8156-3701 & 03:42:22.15 & $-$00:34:59.5 & 0.05273 & 243 & 0 & 0 & 180$\times$9+180$\times$10 & 1 \\
8241-3703 & 08:25:50.69 & +18:10:00.1 & 0.02911 & 132 & 0 & 0 & 180$\times$10 & 1 \\
8241-3704 & 08:26:16.54 & +17:21:44.8 & 0.06617 & 308 & 0 & 0 & 180$\times$8+180$\times$4 & 1 \\
8615-3703 & 21:23:18.34 & +01:15:17.9 & 0.01845 & 83 & 0 & 0 & 180$\times$8 & 1 \\
8615-12702 & 21:20:38.27 & +01:02:50.2 & 0.02095 & 94 & 0 & 0 & 120$\times$5+120$\times$3 & 1 \\
8616-6104 & 21:31:55.33 & +00:12:49.6 & 0.05426 & 250 & 0 & 0 & 180$\times$9 & 1 \\
8618-9102 & 21:17:05.15 & +09:58:20.3 & 0.04334 & 198 & 1 & 0 & 180$\times$10+180$\times$10 & 1 \\
8623-6104 & 20:47:07.44 & +00:18:01.7 & 0.09704 & 461 & 0 & 0 & 180$\times$9 & 3 \\
8623-12702 & 20:40:52.10 & +00:39:10.1 & 0.02691 & 122 & 0 & 0 & 120$\times$3 & 1 \\
8655-3701 & 23:47:00.44 & $-$00:26:50.6 & 0.07149 & 334 & 0 & 0 & 180$\times$9+180$\times$3 & 1 \\
8655-9102 & 23:52:53.26 & $-$00:22:56.8 & 0.04505 & 206 & 2 & 0 & 180$\times$10+180$\times$10 & 1 \\
8728-3701 & 03:50:47.77 & $-$07:01:43.6 & 0.02833 & 128 & 2 & 0 & 120$\times$2+120$\times$4 & 1 \\
8950-12705 & 12:58:55.95 & +27:50:00.4 & 0.02528 & 114 & 2 & 0 & 120$\times$2 & 3 \\
8952-12701 & 13:38:44.12 & +26:19:42.7 & 0.02856 & 129 & 2 & 0 & 120$\times$2 & 3 \\
9194-3702 & 03:08:07.07 & +00:27:22.4 & 0.07454 & 349 & 0 & 1 & 180$\times$10+180$\times$10 & 1 \\
9195-3702 & 01:51:22.27 & +13:03:37.2 & 0.06434 & 299 & 0 & 1 & 120$\times$4+180$\times$4 & 2 \\
9512-3704 & 09:12:36.30 & +00:18:34.2 & 0.05463 & 252 & 2 & 1 & 180$\times$9 & 3 \\
8153-12702 & 02:41:12.96 & $-$00:52:37.3 & 0.03823 & 174 & 1 & 1 & 180$\times$8+180$\times$10 & 1 \\
7975-6104 & 21:39:33.98 & +10:29:00.5 & 0.07920 & 372 & 0 & 1 & 180$\times$10 & 3 \\
8439-6102 & 09:31:06.76 & +49:04:47.1 & 0.03415 & 155 & 2 & 2 & 120$\times$2+180$\times$10 & 1 \\
8250-6104 & 09:21:38.74 & +43:43:34.1 & 0.04008 & 183 & 2 & 2 & 120$\times$7+180$\times$4 & 1 \\
\hline
\end{tabular}
        \begin{tablenotes}
            \item Column (1): the MaNGA Plate-IFU. Columns (2)–(3): equatorial coordinates (J2000). Column (4): redshift from SDSS. Column (5): luminosity distance by adopting cosmology of the Planck Collaboration \cite{PlanckCollaboration2020AA...641A...6P}. Column (6): the status of \HI\ observation in \HI-MaNGA~-- 0: not observed, 1: detection, 2: non-detection. Column (7): the original sample -- 0: ALMaQUEST, 1: the additional sample in ALMaQUEST, 2: Gao \etal\, \cite{Gao2019ApJ...887..172G}. Column (8): the total integration time in unit of second of each cycle with the numbers showing the integration time per cycle and total cycles. The initial and subsequent ones indicate the observation times of 2020 and 2022, respectively. Column (9): the combined quality of FAST \HI\ observation -- 1: detection; 2: non-detection, 3: contaminated by RFI. 
        \end{tablenotes}
    \end{threeparttable}
\end{table*}

\subsection{Sample Definitions}

In the following discussions, we refer to galaxies with 
both published \HI\ and CO spectra as the ``HICO-Spec'' sample. 
This sample comprises galaxies with \HI\ and CO observations (with at least one detection in \HI\ or CO) from the ALMaQUEST and xCOLD GASS samples. After removing 2 duplicate galaxies, the HICO-Spec sample contains 412 galaxies with both \HI\ and CO observation, of which 310 galaxies have both \HI\ and CO detections. In Section~\ref{subsec:rmol-hi}, we investigate the dependence of \HI-H$_2$ conversion on integrated profile shape and asymmetry using this sample. 

\begin{figure}[H]
\centering
\includegraphics[scale=0.5]{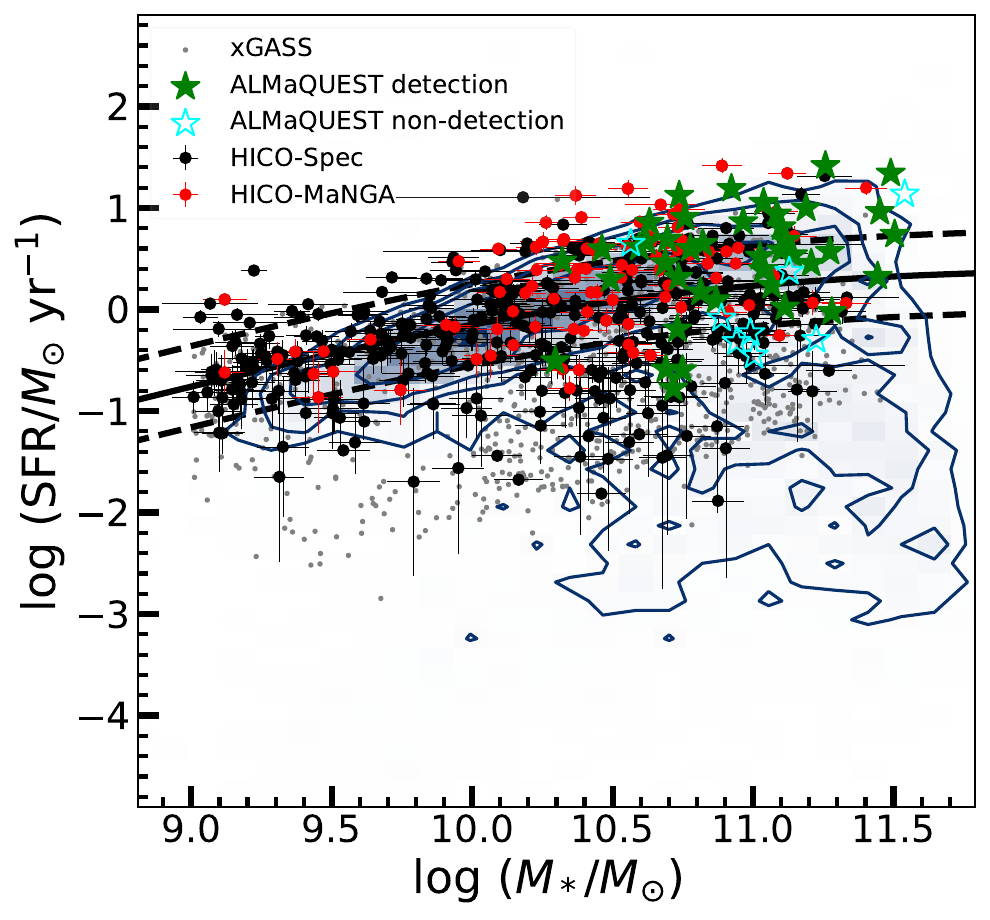}
\caption{
The distribution of galaxies in the HICO-MaNGA (red) and HICO-Spec (black) samples in the space of log SFR vs. log \mstar. The detection and non-detection of the ALMaQUEST galaxies are shown as filled green and open cyan stars, respectively. In comparison, we include MaNGA (blue scale and contours) and xGASS (small grey dots) galaxies in the plot.
The black solid and dashed curves show the star-forming main sequence \cite{Saintonge2022ARA&A..60..319S} and its deviation at $\pm$ 0.4 dex.
}
\label{fig:sfms}
\end{figure}

On the other hand, we refer to the ``HICO-MaNGA'' sample as the cross-match result of the cold gas surveys (ALMaQUEST, xCOLD GASS, ALLSMOG, and JINGLE) and the MaNGA~survey \cite{Sanchez2016RMxAA..52...21S, Sanchez2018RMxAA..54..217S}. During the cross-matching process, we utilize TOPCAT \cite{Taylor2005ASPC..347...29T} and require a maximum projected optical position separation of 3$^{\prime\prime}$ and a velocity separation of 500 km s$^{-1}$. 
We compile the sources in the cold gas catalogs and remove 4 duplicated galaxies, resulting in a total of 390 galaxies with both \HI\ and CO detections. We cross-match these 390 galaxies with the MaNGA~catalog of 10,010 unique galaxies by their positions and redshifts.
This step yields 32 galaxies from xCOLD GASS, 7 galaxies from ALLSMOG, 38 galaxies from ALMaQUEST, and 17 galaxies from the JINGLE survey.
Thus the HICO-MaNGA~sample consists of 94 galaxies with both \HI\ and CO detection as well as optical IFU data from MaNGA. Figure~\ref{fig:sfms} shows the sample distribution in the space of log SFR vs. log \mstar\, for the HICO-Spec and HICO-MaNGA samples.
The sample size of HICO-Spec is slightly larger than that in Saintonge \etal\, \cite{Saintonge2011MNRAS.415...32S}. The galaxy number of the HICO-MaNGA sample is 10\% of the xCOLD GASS survey \cite{Catinella2018MNRAS.476..875C}, but all galaxies in the HICO-MaNGA sample have MaNGA IFU observations. Meanwhile, our sample covers a similar dynamical range in stellar mass and SFR compared to the star-forming galaxies in the MaNGA survey.

\subsection{FAST \HI\ Observations}

In addition to the existing \HI\ observations from GBT and Arecibo, we obtained
FAST \HI\ spectral line observations for the ALMaQUEST and Supplementary Samples. The observation is designed to reach a detection threshold of log \MHI/\mstar\, $\gtrsim$ 1\% following the strategy of GASS and xGASS \cite{Catinella2010MNRAS.403..683C, Catinella2018MNRAS.476..875C}.
In 2020, we conducted the initial \HI\ follow-up observation using FAST (PI: Zheng, project code: PT2020\_0102). We performed additional observations on two galaxies (MaNGA~7977-12705, MaNGA~8077-6104) that were previously detected by \HI-MaNGA~(GBT) to verify the calibration procedure. 
 It should be noted that a portion of the data was affected by strong radio frequency interference (RFI), primarily caused by the refrigerating dewar in the compressor. However, this issue has been partially resolved in 2021 \cite{Xi2022PASA...39...19X}. Taking into account the data obtained in 2020, we have adjusted our observation strategy and proposed a new FAST observation in 2022 led by PI Zheng (project code: PT2022\_0091).

The observations were conducted using the entire wide band of 500 MHz, employing the on-off mode and the 19 beam receiver at the L band (1.05–1.45 GHz). Each beam has a size of 2.9$^{\prime}$ and a sampling time of approximately 0.1 seconds. During each second, the 10 K (high CAL) noise diode is on for the first $\sim$0.1 seconds and off in the remaining 0.9 seconds. The spectrum encompasses 65,536 channels and has a spectral resolution of 7.6 kHz ($\sim$1.6 km s$^{-1}$) for the dual-polarization observations. The separation between the center of beam M01 and one of the outer beams M14 is 11.6$^{\prime}$.
% with a rotation angle set to 0 $^{\circ}$. 

In each cycle of observation, the central beam M01 points on the source and the outer beam M14 points away from the source in the ``source on" mode. Conversely, the beam M14 points on the source and the beam M01 points away from the source in the ``source off" mode. The switching time between source on and source off modes is 30 seconds. Besides the switching or overhead time, the observation beam keeps pointing on the source, effectively doubling the integration time compared to observations without pointing on and off simultaneously. Previous studies \cite{Zheng2020MNRAS.499.3085Z, Yu2022ApJ...934..114Y} have also employed the use of other beams as off-target points to save observation time.

We process the FAST data to obtain the final global spectra. To mitigate the effects of RFI, we remove values with a deviation from the median value larger than 1.5$\sigma$ per channel in each observation cycle. This pre-processing step was necessary for the FAST data of 2020. We then interpolate the remaining FAST raw data using the values of the nearest neighbors to supplement data points. We calibrate the flux intensity using data from the on-off noise diode and applying aperture efficiency correction taking the beam, frequency, and position into consideration \cite{Jiang2019SCPMA..6259502J, Jiang2020RAA....20...64J}. Then the data of the first two polarizations is averaged to derive the preliminary spectrum.

To remove the standing wave\footnote{a kind of periodic sinusoidal feature appeared in the baselines, which is caused by reflected broadband signals in the focal structure of the telescope entering the receiver system with time delay \cite{Briggs1997PASA...14...37B}.}, we fit a sine function after masking the RFI and the velocity range of the potential emission line signal. If the removal of the standing wave did not decrease the noise level of a specific spectrum, we did not adopt the standing wave subtraction. 
Subsequently, a polynomial of order 1-3 was used to fit the spectrum and flatten the baseline. The degree of polynomial order is determined by selecting the degree with the minimum Bayesian Information Criterion (BIC, \cite{Schwarz1978_bic}), which balances the goodness of the fit (reduced chi-square) with the number of parameters in the polynomial fitting. We convert the frequency to heliocentric velocity by using the optical velocity convention and performing the Doppler correction.
For each source, we stack the spectrum of each cycle (at least two cycles for each galaxy by using M01 on and M14 on) by weighting the spectrum according to its noise level \cite{Fabello2011MNRAS.411..993F, Brown2015MNRAS.452.2479B}.

We observed 37 galaxies with a total observation time of 45 hours, which includes the source on, source off, and overhead time. The integration time for each galaxy ranges from 2 minutes to 1 hour. Table~\ref{tab:observation} provides the basic information and observation details for these galaxies. For a RFI masked and baseline subtracted profile, we search for the emission line signal by rebinning the spectra to a channel width of $\sim$ 20 km s$^{-1}$. The noise level of the spectrum ($\sigma_{\rm spec}$) is the standard deviation obtained from the best-fit half-Gaussian fitting to the negative values of all flux intensities within a range of $\pm$ 500 km s$^{-1}$ around the optical central velocity. If the rebinned spectrum contains at least three channels with flux intensities greater than 3$\sigma_{\rm spec}$, it is considered a detection; otherwise, it is classified as a non-detection. The FAST \HI\ spectra quality is listed in Column (9) of Table~\ref{tab:observation}. 

The new FAST observation yields 27 detections, 1 absorption signal, and 2 non-detections, and all thses spectra are public \footnote{\url{https://github.com/NiankunYu/ALMaQUEST_FAST-HI2024}}. Among the 53 galaxies in the ALMaQUEST sample, we have acquired 42 emission line signals, 1 absorption signal, 7 non-detections, and 3 RFI-contaminated ones by combining the data of \HI-MaNGA \cite{Masters2019MNRAS.488.3396M, Stark2021AAS...23752707S} and new FAST observations. There is no useful signal for MaNGA~8082-12704, MaNGA~8623-6104, and MaNGA~7975-6104 due to heavily RFI contamination, and thus were excluded from further discussion. The distribution of ALMaQUEST detection and non-detection of \HI\ signal is shown in Figure~\ref{fig:sfms}. Around half of the non-detections are massive green-valley galaxies with log \mstar $>$ 10.8, which may suggest that galaxy quenching is due to gas depletion.

\begin{figure*}[!htbp]
\centering
\includegraphics[scale=0.4]{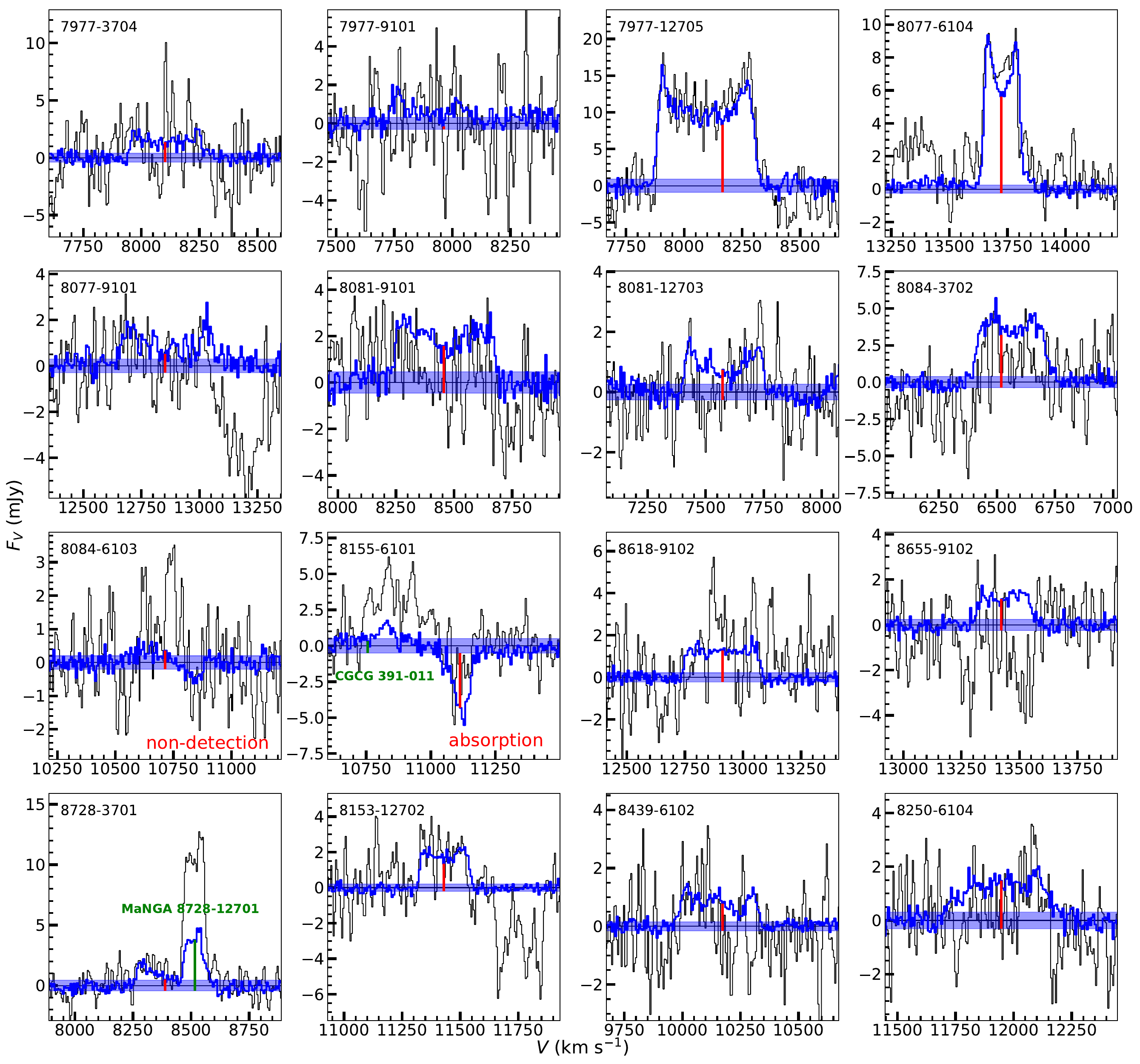}
\caption{The \HI\ spectra of galaxies with both FAST (blue) and \HI-MaNGA (black) observations. In each panel, the MaNGA~IFU plate ID is shown in the upper left corner, the optical central velocity is shown as the red vertical line. Among these profiles, the spectrum of MaNGA~8084-6103 shows no detection above 3-$\sigma$ limit, MaNGA~8155-6101 shows an absorption line signal, and the \HI\ emission of MaNGA 8728-3701 is contaminated by its companion. The optical centers of the companion galaxies of MaNGA~8155-6101 and MaNGA~8728-3701 are shown as green vertical lines.}
\label{fig:hiall}
\end{figure*}

\section{Results and Analysis} 
\label{sec:data}
\subsection{FAST \HI\ Spectra} 
\label{subsec:almaquest}

\begin{figure*}[!htb]
\centering
\includegraphics[scale=0.4]{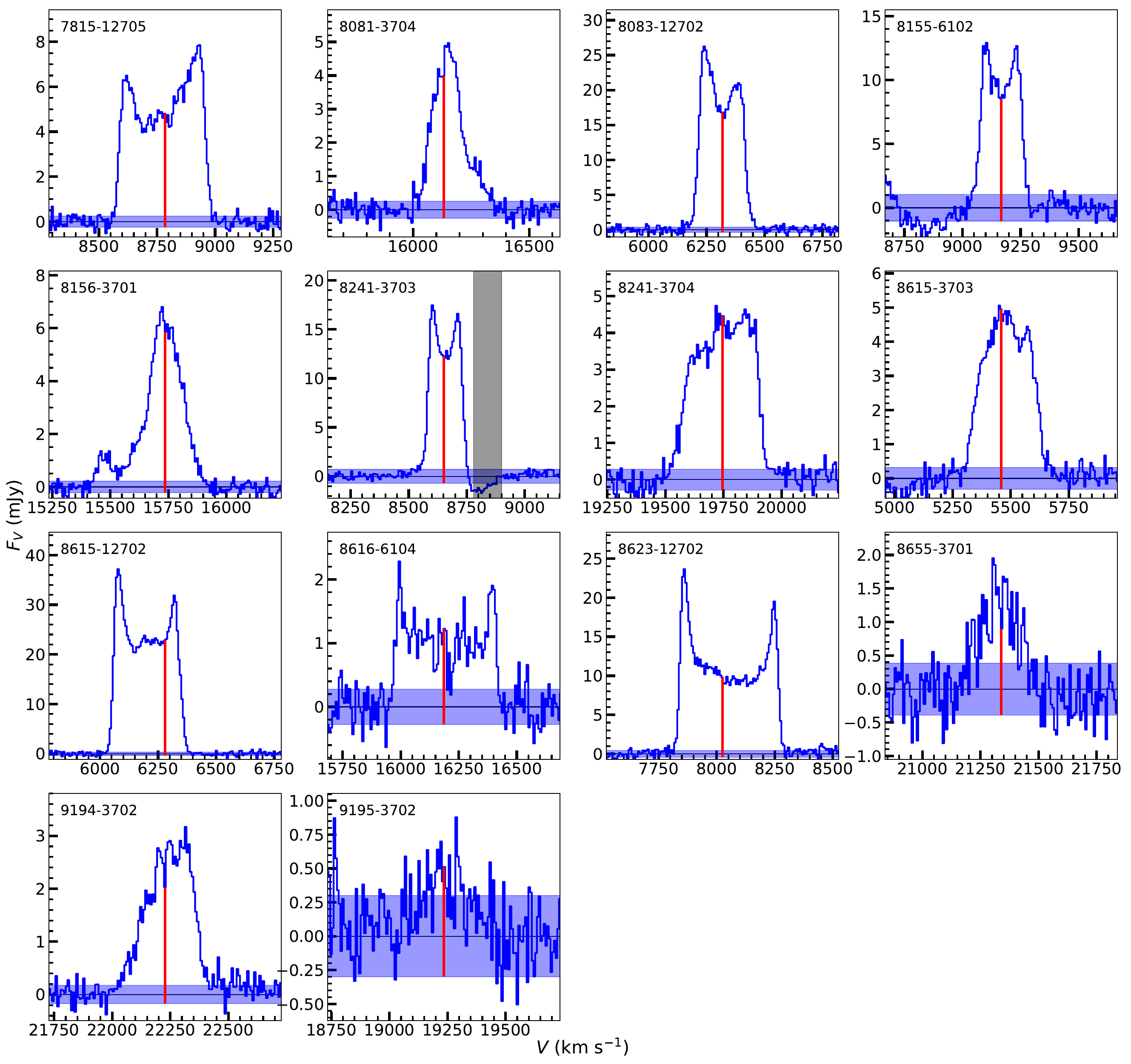}
\caption{The \HI\ spectra of galaxies with only FAST observations.
 In each panel, the MaNGA Plate-IFU is shown in the upper left corner, the optical central velocity is shown as the red vertical line, and the channels contaminated by RFI are shown as grey shaded regions. The spectrum of MaNGA 9195-3702 is identified as a non-detection. }
\label{fig:hiall-only}
\end{figure*}

The final results of FAST observations are shown as blue profiles in Figure~\ref{fig:hiall} and \ref{fig:hiall-only}, which includes emission, absorption, and non-detection. To align with the spectra obtained from Arecibo and GBT, we rebin the spectra to a channel width of $\sim$ 6 km s$^{-1}$. The measurements of spectral noise level $\sigma$ (in units of mJy per channel) follows the same approach as $\sigma_{\rm spec}$, but with a narrower channel width $\Delta V$ $\sim$ 6.4 km s$^{-1}$. The black profiles in Figure~\ref{fig:hiall} are observations from the \HI-MaNGA~survey. The comparison of the blue and black global profiles in Figure~\ref{fig:hiall} demonstrates the excellent consistency between the FAST and archival (Arecibo and GBT) spectra, but the new data exhibit a relatively lower noise level.

While the \HI-MaNGA~data indicate a non-detection for MaNGA~8155-6101, the FAST data reveals a high S/N \HI\ absorption signal by reducing the noise level by half, which is submerged in the noise in the GBT spectrum (we will return to a discussion of this absorption signal in Section~\ref{subsec:absorption}). The discrepancy in spectra between FAST and GBT demonstrates the capability of FAST in detecting new signals. Regarding MaNGA~8084-6103, the new FAST observation further constrains the upper limit of the \HI\ mass to $M_{\rm H\ I, lim}= 10^{8.5}$ \msun, which is 1.1 dex deeper than that of \HI-MaNGA. 
The upper limit of \HI\ mass is calculated by assuming a line width of 200 \kms\ and a 3$\sigma$ detection following \HI-MaNGA~\cite{Masters2019MNRAS.488.3396M, Stark2021AAS...23752707S}. So the upper limits for the \HI\ mass is
% \begin{equation}
\begin{multline}
M_{\rm H\ I, lim}/M_{\odot}  =  2.356\times10^5\times(\frac{D_L}{\rm Mpc})^2\times \\
% (\frac{\sqrt{200\times (\Delta V)\times 3/1000}}{\rm Jy\ km\ s^{-1}})
(\frac{\sqrt{200\times (\Delta V/ {\rm km\ s}^{-1})}\times 3 \sigma/(1000{\rm mJy})}{\rm Jy\ km\ s^{-1}}),
\end{multline}
and the \HI\ mass upper limit of 7 non-detections are shown in Table~\ref{tab:non-detection}. Additionally, the \HI\ profile of MaNGA~8728-3701 from \HI-MaNGA is classified as a non-detection, while the strong \HI\ emission at 8400-8600 km s$^{-1}$ originates from its companion, MaNGA~8728-12701. The new FAST data provides further confirmation that the weak emission around 8300 km s$^{-1}$ may be from MaNGA~8728-3701 and give a conservative \HI\ mass upper limit of $10^{9.3}$ \msun. Due to the inclusion of the emission at 8200--8600 km\,s$^{-1}$, our \HI\ mass upper limit exceeds that of \HI-MaNGA.

% \clearpage
\begin{table*}[htbp]
\centering
\footnotesize
    \begin{threeparttable}
        \caption{Parameters Derived from the \HI\ Spectra of the ALMaQUEST Sample}
        \label{tab:co-add}
        \centering
        \begin{tabular}{lrrrrrrrrrr}
\\
\hline%
\hline%
Plate-IFU & $V_c$ & $F$ & $V_{85}$ & $A_F$ & $A_C$ & $C_V$ & $K$ & S/N & $\sigma$ & log \MHI \\
 & (\kms) & (Jy \kms) & (\kms) & & & & & & (mJy) & (\msun) \\
(1) & (2) & (3) & (4) & (5) & (6) & (7) & (8) & (9) & (10) & (11)\\
\hline
7815-12705 & 8785 $\pm$ 3 & 2.10 $\pm$ 0.33 & 329 $\pm$ 20 & 1.09 $\pm$ 0.08 & 1.30 $\pm$ 0.13 & 2.95 $\pm$ 0.21 & $-$0.04 $\pm$ 0.02 & 108.9 & 0.2 & 9.9 $\pm$ 0.1 \\
7977-3704 & 8114 $\pm$ 6 & 0.50 $\pm$ 0.08 & 285 $\pm$ 17 & 1.04 $\pm$ 0.08 & 1.03 $\pm$ 0.12 & 2.93 $\pm$ 0.23 & $-$0.04 $\pm$ 0.02 & 18.2 & 0.4 & 9.2 $\pm$ 0.1 \\
7977-9101 & 7888 $\pm$ 9 & 0.29 $\pm$ 0.05 & 348 $\pm$ 26 & 1.00 $\pm$ 0.09 & 1.45 $\pm$ 0.17 & 2.49 $\pm$ 0.22 & $-$0.07 $\pm$ 0.02 & 13.1 & 0.3 & 9.0 $\pm$ 0.1 \\
7977-12705 & 8094 $\pm$ 3 & 4.61 $\pm$ 0.73 & 371 $\pm$ 22 & 1.00 $\pm$ 0.07 & 1.00 $\pm$ 0.10 & 3.06 $\pm$ 0.22 & $-$0.03 $\pm$ 0.02 & 66.1 & 0.9 & 10.2 $\pm$ 0.1 \\
8077-6104 & 13724 $\pm$ 3 & 1.27 $\pm$ 0.20 & 153 $\pm$ 9 & 1.05 $\pm$ 0.08 & 1.08 $\pm$ 0.11 & 2.92 $\pm$ 0.21 & $-$0.03 $\pm$ 0.02 & 73.3 & 0.2 & 10.1 $\pm$ 0.1 \\
8077-9101 & 12849 $\pm$ 7 & 0.50 $\pm$ 0.08 & 438 $\pm$ 33 & 1.00 $\pm$ 0.08 & 1.25 $\pm$ 0.15 & 2.62 $\pm$ 0.25 & $-$0.05 $\pm$ 0.03 & 20.8 & 0.3 & 9.6 $\pm$ 0.1 \\
8078-6103 & 8516 $\pm$ 10 & 2.03 $\pm$ 0.33 & 276 $\pm$ 24 & 1.14 $\pm$ 0.13 & 1.04 $\pm$ 0.13 & 4.22 $\pm$ 0.39 & 0.04 $\pm$ 0.02 & 10.4 & 3.3 & 9.9 $\pm$ 0.1 \\
8078-12701 & 7995 $\pm$ 11 & 3.41 $\pm$ 0.53 & 408 $\pm$ 25 & 1.08 $\pm$ 0.10 & 1.36 $\pm$ 0.23 & 2.33 $\pm$ 0.20 & $-$0.10 $\pm$ 0.02 & 10.6 & 4.7 & 10.0 $\pm$ 0.1 \\
8081-3704 & 16160 $\pm$ 3 & 0.75 $\pm$ 0.12 & 190 $\pm$ 12 & 1.19 $\pm$ 0.09 & 1.32 $\pm$ 0.15 & 4.83 $\pm$ 0.34 & 0.09 $\pm$ 0.02 & 43.2 & 0.3 & 10.0 $\pm$ 0.1 \\
8081-9101 & 8452 $\pm$ 6 & 0.91 $\pm$ 0.14 & 393 $\pm$ 25 & 1.12 $\pm$ 0.08 & 1.10 $\pm$ 0.12 & 2.81 $\pm$ 0.21 & $-$0.04 $\pm$ 0.02 & 24.2 & 0.5 & 9.5 $\pm$ 0.1 \\
8081-9102 & 10131 $\pm$ 8 & 2.40 $\pm$ 0.38 & 390 $\pm$ 24 & 1.02 $\pm$ 0.08 & 1.12 $\pm$ 0.13 & 2.69 $\pm$ 0.21 & $-$0.05 $\pm$ 0.02 & 18.5 & 1.8 & 10.1 $\pm$ 0.1 \\
8081-12703 & 7578 $\pm$ 6 & 0.33 $\pm$ 0.05 & 310 $\pm$ 19 & 1.00 $\pm$ 0.08 & 1.02 $\pm$ 0.14 & 2.49 $\pm$ 0.20 & $-$0.08 $\pm$ 0.02 & 17.4 & 0.3 & 9.0 $\pm$ 0.1 \\
8082-6103 & 7207 $\pm$ 5 & 1.42 $\pm$ 0.22 & 268 $\pm$ 19 & 1.30 $\pm$ 0.11 & 1.74 $\pm$ 0.20 & 4.24 $\pm$ 0.39 & 0.07 $\pm$ 0.02 & 14.5 & 1.8 & 9.6 $\pm$ 0.1 \\
8082-12701 & 8034 $\pm$ 9 & 2.88 $\pm$ 0.45 & 302 $\pm$ 21 & 1.10 $\pm$ 0.09 & 1.17 $\pm$ 0.15 & 2.80 $\pm$ 0.25 & $-$0.05 $\pm$ 0.02 & 11.5 & 4.1 & 10.0 $\pm$ 0.1 \\
8083-6101 & 7938 $\pm$ 6 & 3.72 $\pm$ 0.59 & 382 $\pm$ 26 & 1.01 $\pm$ 0.08 & 1.38 $\pm$ 0.15 & 2.96 $\pm$ 0.23 & $-$0.03 $\pm$ 0.02 & 24.6 & 2.3 & 10.1 $\pm$ 0.1 \\
8083-9101 & 11445 $\pm$ 13 & 1.54 $\pm$ 0.24 & 357 $\pm$ 30 & 1.16 $\pm$ 0.15 & 1.00 $\pm$ 0.15 & 2.99 $\pm$ 0.39 & $-$0.05 $\pm$ 0.03 & 11.9 & 1.8 & 10.0 $\pm$ 0.1 \\
8083-12702 & 6312 $\pm$ 3 & 4.36 $\pm$ 0.69 & 183 $\pm$ 11 & 1.06 $\pm$ 0.07 & 1.17 $\pm$ 0.12 & 2.95 $\pm$ 0.21 & $-$0.03 $\pm$ 0.02 & 168.2 & 0.4 & 10.0 $\pm$ 0.1 \\
8084-3702 & 6562 $\pm$ 4 & 1.18 $\pm$ 0.19 & 261 $\pm$ 16 & 1.05 $\pm$ 0.08 & 1.06 $\pm$ 0.11 & 3.08 $\pm$ 0.22 & $-$0.02 $\pm$ 0.02 & 42.3 & 0.4 & 9.4 $\pm$ 0.1 \\
8084-12705 & 7543 $\pm$ 6 & 2.88 $\pm$ 0.45 & 405 $\pm$ 26 & 1.18 $\pm$ 0.09 & 1.32 $\pm$ 0.14 & 3.00 $\pm$ 0.23 & $-$0.01 $\pm$ 0.02 & 24.4 & 1.7 & 9.9 $\pm$ 0.1 \\
8086-9101 & 11979 $\pm$ 13 & 1.18 $\pm$ 0.18 & 365 $\pm$ 31 & 1.00 $\pm$ 0.09 & 1.35 $\pm$ 0.25 & 4.03 $\pm$ 0.36 & 0.02 $\pm$ 0.02 & 12.5 & 1.2 & 9.9 $\pm$ 0.1 \\
8155-6102 & 9165 $\pm$ 4 & 2.05 $\pm$ 0.32 & 167 $\pm$ 10 & 1.00 $\pm$ 0.07 & 1.01 $\pm$ 0.11 & 3.02 $\pm$ 0.22 & $-$0.03 $\pm$ 0.02 & 33.7 & 1.0 & 10.0 $\pm$ 0.1 \\
8156-3701 & 15712 $\pm$ 4 & 1.22 $\pm$ 0.19 & 259 $\pm$ 16 & 1.34 $\pm$ 0.10 & 1.72 $\pm$ 0.18 & 5.22 $\pm$ 0.37 & 0.10 $\pm$ 0.02 & 67.6 & 0.2 & 10.2 $\pm$ 0.1 \\
8241-3703 & 8655 $\pm$ 3 & 2.25 $\pm$ 0.36 & 135 $\pm$ 8 & 1.07 $\pm$ 0.08 & 1.04 $\pm$ 0.11 & 2.97 $\pm$ 0.21 & $-$0.03 $\pm$ 0.02 & 54.7 & 0.7 & 10.0 $\pm$ 0.1 \\
8241-3704 & 19748 $\pm$ 4 & 1.36 $\pm$ 0.22 & 293 $\pm$ 18 & 1.10 $\pm$ 0.08 & 1.15 $\pm$ 0.12 & 3.66 $\pm$ 0.26 & 0.01 $\pm$ 0.02 & 56.6 & 0.3 & 10.5 $\pm$ 0.1 \\
8450-6102 & 12533 $\pm$ 7 & 0.78 $\pm$ 0.14 & 101 $\pm$ 13 & 1.09 $\pm$ 0.19 & 1.29 $\pm$ 0.33 & 5.27 $\pm$ 0.90 & 0.08 $\pm$ 0.03 & 8.6 & 2.3 & 9.8 $\pm$ 0.1 \\
8615-3703 & 5483 $\pm$ 4 & 1.21 $\pm$ 0.19 & 248 $\pm$ 15 & 1.02 $\pm$ 0.07 & 1.01 $\pm$ 0.10 & 3.92 $\pm$ 0.28 & 0.03 $\pm$ 0.02 & 50.8 & 0.3 & 9.3 $\pm$ 0.1 \\
8615-9101 & 9961 $\pm$ 6 & 1.88 $\pm$ 0.29 & 402 $\pm$ 24 & 1.04 $\pm$ 0.08 & 1.06 $\pm$ 0.12 & 2.93 $\pm$ 0.22 & $-$0.05 $\pm$ 0.02 & 20.0 & 1.3 & 10.0 $\pm$ 0.1 \\
8615-12702 & 6198 $\pm$ 3 & 7.55 $\pm$ 1.19 & 257 $\pm$ 15 & 1.01 $\pm$ 0.07 & 1.02 $\pm$ 0.10 & 3.07 $\pm$ 0.22 & $-$0.04 $\pm$ 0.02 & 303.8 & 0.4 & 10.2 $\pm$ 0.1 \\
8616-6104 & 16189 $\pm$ 7 & 0.51 $\pm$ 0.08 & 408 $\pm$ 25 & 1.05 $\pm$ 0.08 & 1.12 $\pm$ 0.13 & 2.84 $\pm$ 0.21 & $-$0.04 $\pm$ 0.02 & 21.8 & 0.3 & 9.9 $\pm$ 0.1 \\
8616-9102 & 9048 $\pm$ 4 & 3.23 $\pm$ 0.51 & 212 $\pm$ 13 & 1.01 $\pm$ 0.07 & 1.00 $\pm$ 0.10 & 3.03 $\pm$ 0.22 & $-$0.03 $\pm$ 0.02 & 36.0 & 1.6 & 10.2 $\pm$ 0.1 \\
8616-12702 & 9131 $\pm$ 6 & 2.96 $\pm$ 0.47 & 450 $\pm$ 28 & 1.22 $\pm$ 0.09 & 1.83 $\pm$ 0.20 & 2.84 $\pm$ 0.23 & $-$0.02 $\pm$ 0.02 & 26.5 & 1.6 & 10.1 $\pm$ 0.1 \\
8618-9102 & 12913 $\pm$ 5 & 0.40 $\pm$ 0.06 & 273 $\pm$ 16 & 1.00 $\pm$ 0.07 & 1.00 $\pm$ 0.10 & 3.44 $\pm$ 0.25 & $-$0.01 $\pm$ 0.02 & 26.4 & 0.2 & 9.6 $\pm$ 0.1 \\
8623-12702 & 8042 $\pm$ 3 & 5.34 $\pm$ 0.84 & 391 $\pm$ 23 & 1.07 $\pm$ 0.07 & 1.25 $\pm$ 0.13 & 2.87 $\pm$ 0.20 & $-$0.06 $\pm$ 0.02 & 156.4 & 0.4 & 10.3 $\pm$ 0.1 \\
8655-3701 & 21331 $\pm$ 7 & 0.27 $\pm$ 0.04 & 216 $\pm$ 16 & 1.00 $\pm$ 0.09 & 1.00 $\pm$ 0.12 & 4.59 $\pm$ 0.41 & 0.08 $\pm$ 0.02 & 11.1 & 0.4 & 9.8 $\pm$ 0.1 \\
8655-9102 & 13437 $\pm$ 5 & 0.30 $\pm$ 0.05 & 207 $\pm$ 13 & 1.06 $\pm$ 0.08 & 1.15 $\pm$ 0.13 & 3.12 $\pm$ 0.24 & $-$0.02 $\pm$ 0.02 & 18.6 & 0.2 & 9.5 $\pm$ 0.1 \\
8655-12705 & 13413 $\pm$ 9 & 1.07 $\pm$ 0.17 & 412 $\pm$ 28 & 1.29 $\pm$ 0.11 & 1.30 $\pm$ 0.17 & 3.31 $\pm$ 0.29 & 0.05 $\pm$ 0.02 & 13.3 & 1.2 & 10.0 $\pm$ 0.1 \\
%8728-3701 & 8453 $\pm$ 6 & 0.55 $\pm$ 0.09 & 257 $\pm$ 17 & 1.81 $\pm$ 0.16 & 4.83 $\pm$ 0.69 & 2.84 $\pm$ 0.33 & $-$0.02 $\pm$ 0.03 & 18.2 & 0.4 & 9.3 $\pm$ 0.1 \\
8952-6104 & 8460 $\pm$ 4 & 1.70 $\pm$ 0.27 & 62 $\pm$ 5 & 1.18 $\pm$ 0.16 & 1.30 $\pm$ 0.19 & 3.67 $\pm$ 0.46 & 0.03 $\pm$ 0.02 & 14.3 & 2.9 & 9.8 $\pm$ 0.1 \\
9194-3702 & 22245 $\pm$ 4 & 0.70 $\pm$ 0.11 & 264 $\pm$ 17 & 1.14 $\pm$ 0.08 & 1.22 $\pm$ 0.13 & 3.92 $\pm$ 0.30 & 0.05 $\pm$ 0.02 & 48.2 & 0.2 & 10.3 $\pm$ 0.1 \\
8153-12702 & 11435 $\pm$ 4 & 0.40 $\pm$ 0.06 & 184 $\pm$ 11 & 1.01 $\pm$ 0.07 & 1.04 $\pm$ 0.11 & 3.13 $\pm$ 0.23 & $-$0.02 $\pm$ 0.02 & 33.1 & 0.2 & 9.4 $\pm$ 0.1 \\
8439-6102 & 10145 $\pm$ 5 & 0.30 $\pm$ 0.05 & 306 $\pm$ 18 & 1.05 $\pm$ 0.08 & 1.39 $\pm$ 0.15 & 3.98 $\pm$ 0.31 & $-$0.00 $\pm$ 0.02 & 26.9 & 0.2 & 9.2 $\pm$ 0.1 \\
8250-6104 & 11957 $\pm$ 7 & 0.59 $\pm$ 0.09 & 376 $\pm$ 24 & 1.00 $\pm$ 0.08 & 1.04 $\pm$ 0.11 & 3.56 $\pm$ 0.27 & 0.01 $\pm$ 0.02 & 22.9 & 0.3 & 9.7 $\pm$ 0.1 \\
\hline
\end{tabular}
        \begin{tablenotes}
            \item Column (1): the MaNGA Plate-IFU. Column (2): Flux intensity-weighted central velocity. Column (3): Total global flux of the \HI\ line. Column (4): Velocity width measured at 85\% of the total line flux. Column (5): Global flux asymmetry. Column (6): Flux distribution asymmetry. Column (7): Profile concentration. Column (8): Profile shape. Column (9): S/N of the profile. Column (10): Noise level of the profile at a channel width of $\sim$6.4 km\,s$^{-1}$. Column (11): \HI\ mass and its uncertainty; for a distance uncertainty of 10\% and a flux uncertainty of 15\%, the typical uncertainty of log~\MHI\ is 0.1 dex.
        \end{tablenotes}
    \end{threeparttable}
\end{table*}

\subsection{Measurements of Global Profiles} 
\label{subsec:measure}
The \HI\ emission line spectra of the new FAST supplements and xCOLD GASS datasets are uniformly measured using the curve-of-growth method \cite{Yu2020ApJ...898..102Y, Yu2022ApJS..261...21Y}. This method defines the profile center as the flux-intensity weighted velocity, and constructs the curve of growth by accumulating flux intensities from the center to two sides, then quantifies the total flux, line width, asymmetry, and profile shape. We remove two galaxies: MaNGA 8728-3701 and MaNGA 8155-6101 due to contamination of companions and absorption signals in section~\ref{sec:HIH2}.

The total flux $F$ (in units of Jy \kms) is determined as the median value of the flat part of the growth curve. The line width is defined as the velocity width that encloses a characteristic fraction of the total flux. For instance, line widths $V_{25}$ and $V_{85}$ correspond to the velocity widths enclosing 25\% and 85\% of the total flux, respectively. The profile asymmetry is quantified as the ratio of the integrated fluxes for two sides ($A_F\geq$1.0) and the ratio of the slopes of the rising part of the curve of growth for two sides ($A_C\geq$1.0). We quantify the emission line profile shape as the degree of the concentration of the line profile $C_V$ and the integrated area between the normalized curve of growth and the diagonal line of unity $K$. With the increasing of $C_V$ or $K$, the profile transits from a double-horned profile to a single-peaked profile. The S/N of a spectrum is calculated as
\begin{equation}
S/N = \frac{1000F}{\sigma \Delta V\sqrt{2N}}.
\end{equation}
We apply a second-order polynomial correction to account for systematic bias in the total flux, line width, profile asymmetry, and profile shape when the spectrum has a S/N below 40 following Yu \etal\, \cite{Yu2022ApJS..261...21Y}. The \HI\ mass is calculated as \MHI\ = $2.356\times10^5D_L^2F$ \msun\ \cite{Roberts1962AJ.....67..437R}, where $D_L$ is the luminosity distance with a unit of Mpc. The FAST beam roughly corresponds to a physical scale of 200 kpc, which is enough to cover all \HI\ emissions within the host galaxy, thus the aperture correction is not necessary for our targets.
The final measured parameters of all detections obtained from our \HI\ spectra are presented in Table~\ref{tab:co-add}.

% \end{equation}
The \HI\ mass upper limits of the ALMaQUEST sample cover a range of $10^{8.2}$ to $10^{9.8}$ \msun\ (Table~\ref{tab:non-detection}), which correspond to a gas fraction log \MHI/\mstar\ of 0.5--10\% with a median value of 2\%. The two \HI\ non-detections of FAST further constrain the atomic-to-stellar mass ratio to 0.5\%, which is consistent with our expectations of 1\%. The noise level of our new FAST observation is $\sim$ 3--7 times better than that from the GBT or Arecibo, which is due to the large dish size thus high sensitivity of FAST compared to GBT; and significantly longer integration times relative to the ALFALFA survey.

%%%%%%%%%%%%%%%%%%%%%%%%%%%%%%%%%%%%%%%%%%%%%%%%%%%%%%%%%%%%%%%%%%%%%%%%%%%%%%%%%%%%
% %%%%%%%%%%%%%%%%%%%%%%%%%%%% %%%%%%%%%%%%%%%%%%%%%%%%%%% ynk: put table 2, 3 here
%%%%%%%%%%%%%%%%%%%%%%%%%%%%%%%%%%%%%%%%%%%%%%%%%%%%%%%%%%%%%%%%%%%%%%%%%%%%%%%%%%%%

\begin{table}[H]
\footnotesize
\begin{threeparttable}\caption{Upper Limits of \HI\ Mass}\label{tab:non-detection}
\doublerulesep 0.1pt \tabcolsep 13pt %space between two columns. ????????§Þ???
\begin{tabular}{lrrc}
\toprule
Plate-IFU & $\sigma$ & log $M_{\rm H\ I, lim}$ & Notes \\
 & (mJy) & ($M_{\odot}$) & \\
(1) & (2) & (3) & (4) \\
\hline
7977-3703\tnote{1} & 2.6 & 9.0 & 2 \\
8081-6102 & 1.2 & 8.9 & 2 \\
8084-6103\tnote{1} & 1.4 & 9.3 & 2 \\
 & 0.2 & 8.2 & 1 \\
8728-3701\tnote{2} & 1.3 & 8.9 & 2 \\
 & 0.4 & 9.3 & 1 \\
8950-12705 & 2.6 & 8.9 & 2 \\
8952-12701 & 2.6 & 9.0 & 2 \\
9195-3702 & 0.3 & 8.8 & 1 \\
9512-3704 & 3.9 & 9.8 & 2 \\
\bottomrule
\end{tabular}
Column (1): the MaNGA Plate-IFU. Columns (2): noise level of the \HI\ line profile. Column (3): upper limit of the \HI\ mass, assuming a 3$\sigma$ detection with a line width of 200 km\,s$^{-1}$. Column (4): the \HI\ spectrum is from 1: FAST and 2: \HI-MaNGA observation.
\begin{tablenotes}
\item[1] The \HI\ spectra of MaNGA 7977-3703 and MaNGA 8084-6103 are classified as detections in \HI-MaNGA.
\item[2] The \HI\ emission of MaNGA 8728-3701 is contaminated with that of its companion MaNGA 8728-12701, thus we calculate the upper limit of their total \HI\ mass.
\end{tablenotes}
\end{threeparttable}
\end{table}

\subsection{\HI\ Absorption in MaNGA~8155-6101}
\label{subsec:absorption}
\begin{figure*}[!htb]
\centering
\includegraphics[scale=0.4]{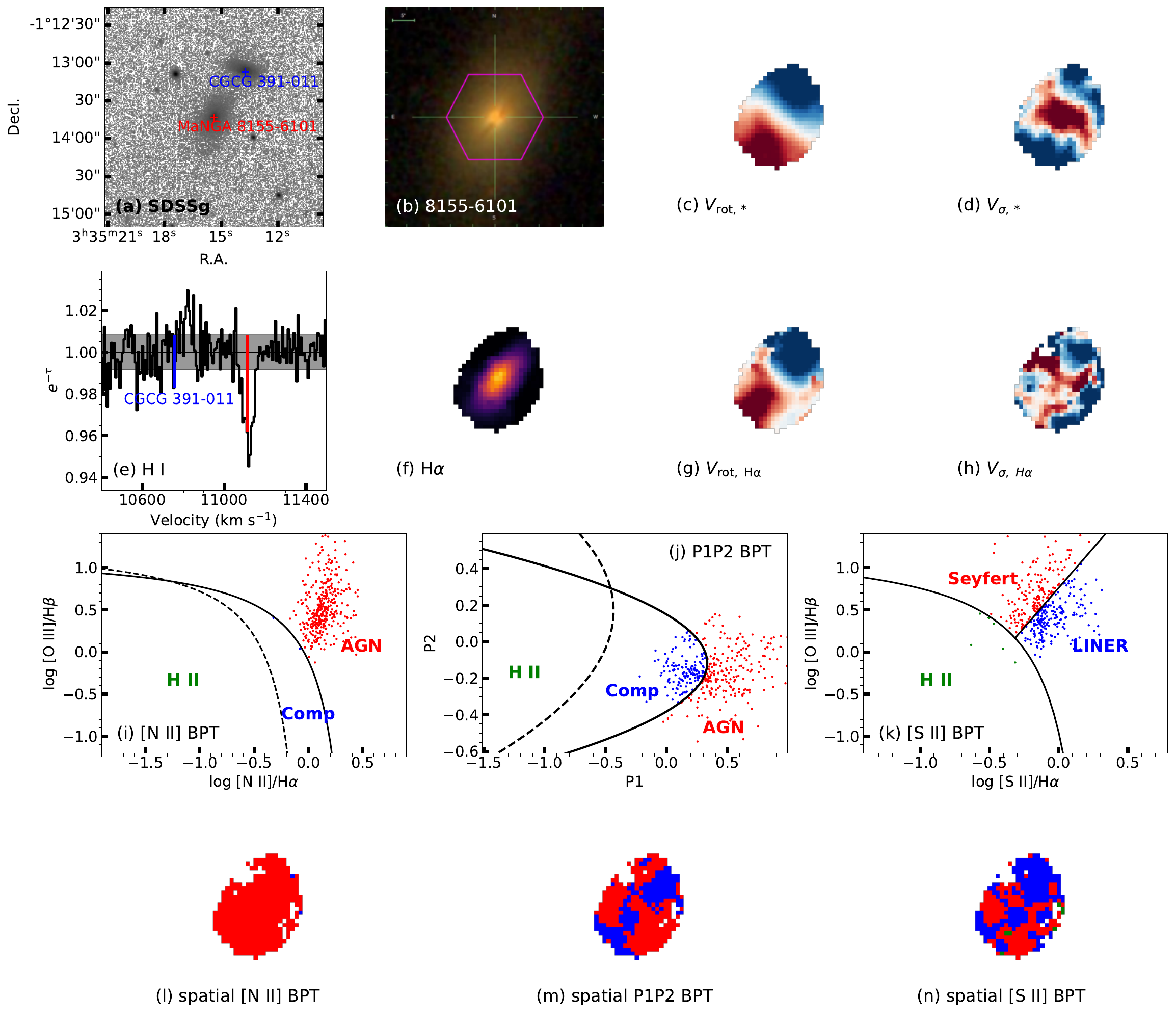}
\caption{The physical properties of MaNGA~8155-6101, which is an interacting galaxy with \HI\ absorption signal. Panel (a) shows the SDSS $g$-band image of MaNGA~8155-6101 with its companion CGCG~391-011, and the field of view is 2.9$^{\prime}\times 2.9 ^{\prime}$ (same as FAST beam size). Panel (b) is the SDSS $gri$ color composite image of MaNGA~8155-6101. Panel (c) and (d) are the velocity field and velocity dispersion of the stellar component for MaNGA~8155-6101. Panel (e) is $e^{-\tau}$ as a function of velocity for MaNGA~8155-6101 and its companion, where $\tau$ is the optical depth of the absorbing gas following Allison \etal\, \cite{Allison2013MNRAS.430..157A}. Panels (f), (g), and (h) are the H$\alpha$ image, velcoity field, and velocity dispersion field of MaNGA~8155-6101. Panels (i)--(n) represent the classification and spatial distribution of three ionization states distinguished by three different BPT diagrams.}
\label{fig:absorption}
\end{figure*}

%%%%%%%%%%%%%%%%%%%%%%%%%%%%%%%%%%%%%%%%%%%%%
Remarkably, MaNGA~8155-6101 has a $\sim$ 6$\sigma$ \HI\ absorption signal (Figure~\ref{fig:absorption}). MaNGA~8155-6101 is a merging galaxy system with a companion CGCG~391-011 at a velocity difference of 360 km s$^{-1}$ and a projected physical separation of $\sim$ 30 kpc (44$^{\prime\prime}$). To investigate the absorption feature, we follow Wolfe \& Burbidge \cite{Wolfe1975ApJ...200..548W} and Allison \etal\, \cite{Allison2013MNRAS.430..157A}, and the \HI\ column density from the absorption line signal is determined as 
\begin{equation}
\begin{split}
    N_{\HI} &= 1.823\times 10^{18} T_{\rm spin} \int \tau(V)dV,\\
            & \approx 1.823\times 10^{18} \frac{T_{\rm spin}}{f} \int \tau_{\rm obs}(V)dV,\\
\end{split}
\end{equation}
where $\tau(V)$ is the optical depth at a given velocity $V$, $\tau_{\rm obs}(V)$ is the observed optical depth (panel e of Figure~\ref{fig:absorption}), $T_{\rm spin}$ is the mean harmonic spin temperature of the gas, and $f$ is the covering factor of \HI\ absorption. The \HI\ spin temperature ranges from 10 to 1000 K by analysing the \HI\ emission and
absorption profiles towards individual
clouds located in front
of Galactic and extra-galactic radio continuum sources \cite{Mebold1982A&A...115..223M, Heiles2003ApJ...586.1067H, Murray2015ApJ...804...89M, Sofue2017MNRAS.468.4030S}. The covering factor $f$ varies from 0.1 to 1 \cite{Curran2005MNRAS.356.1509C} for galaxies with $z\lesssim$1, and we often assume it to be 1 due to lack of information \cite{Morganti2018A&ARv..26....4M}.
We adopt $T_{\rm spin}$ = 100 K and $f$ = 1 following Carilli \etal\, \cite{Carilli1992ApJ...400L..13C}. 
This results in a \HI\ column density of $5.5 \times 10^{20}$ cm$^{-2}$, which is within the typical range of $10^{19}$--$10^{21}$ cm$^{-2}$ \cite{Murray2015ApJ...804...89M}. But given the uncertainty in the spin temperature and the covering fraction, the uncertainty of \HI\ column density could be one or two orders of magnitude. The center of the absorption signal is precisely consistent with the optical central velocity of MaNGA~8155-6101, while the emission line signal on the blue-shifted side suggests a potential association with the neighboring companion, CGCG 391-011.

Based on spatially resolved stellar kinematics, MaNGA~8155-6101 has a regular rotation pattern and high velocity dispersion within one effective radius (Panel c of Figure~\ref{fig:absorption}). However, the ionized gas kinematic traced by H$\alpha$ emission shows significant perturbation. The CO emission line profile is quite extended with a line width $\approx$ 600 km s$^{-1}$, even though the S/N is low \cite{Lin2020ApJ...903..145L}. 
The broad CO line width is rare considering there are only two galaxies with CO line widths larger than 600 km s$^{-1}$ among 333 detections in the xCOLD GASS sample \cite{Saintonge2017ApJS..233...22S}, which may suggest that the CO in two interacting galaxies are perturbed and mixed. Therefore, the molecular gas in MaNGA 8155-6101 are potentially perturbed, either by AGN or galaxy interaction.
% The broad CO line width and \HI\ absorption suggest that the cold gas is perturbed due to galaxy interaction. 
Furthermore, its SFR, specific star formation rate (sSFR), and star formation efficiency are typical for a massive quenched galaxy with the molecular gas fraction of 1\% \cite{Lin2020ApJ...903..145L}. Thorp \etal\, \cite{Thorp2022MNRAS.516.1462T} found that the central bursts of star formation in mergers require centralized enhancements in gas fraction, thus the broad CO line width and normal SFR in MaNGA~8155-6101 may suggest that gas is distributed extensively thus failed to fuel central star burst activities.

\begin{figure*}[!htb]
\centering
\includegraphics[scale=0.4]{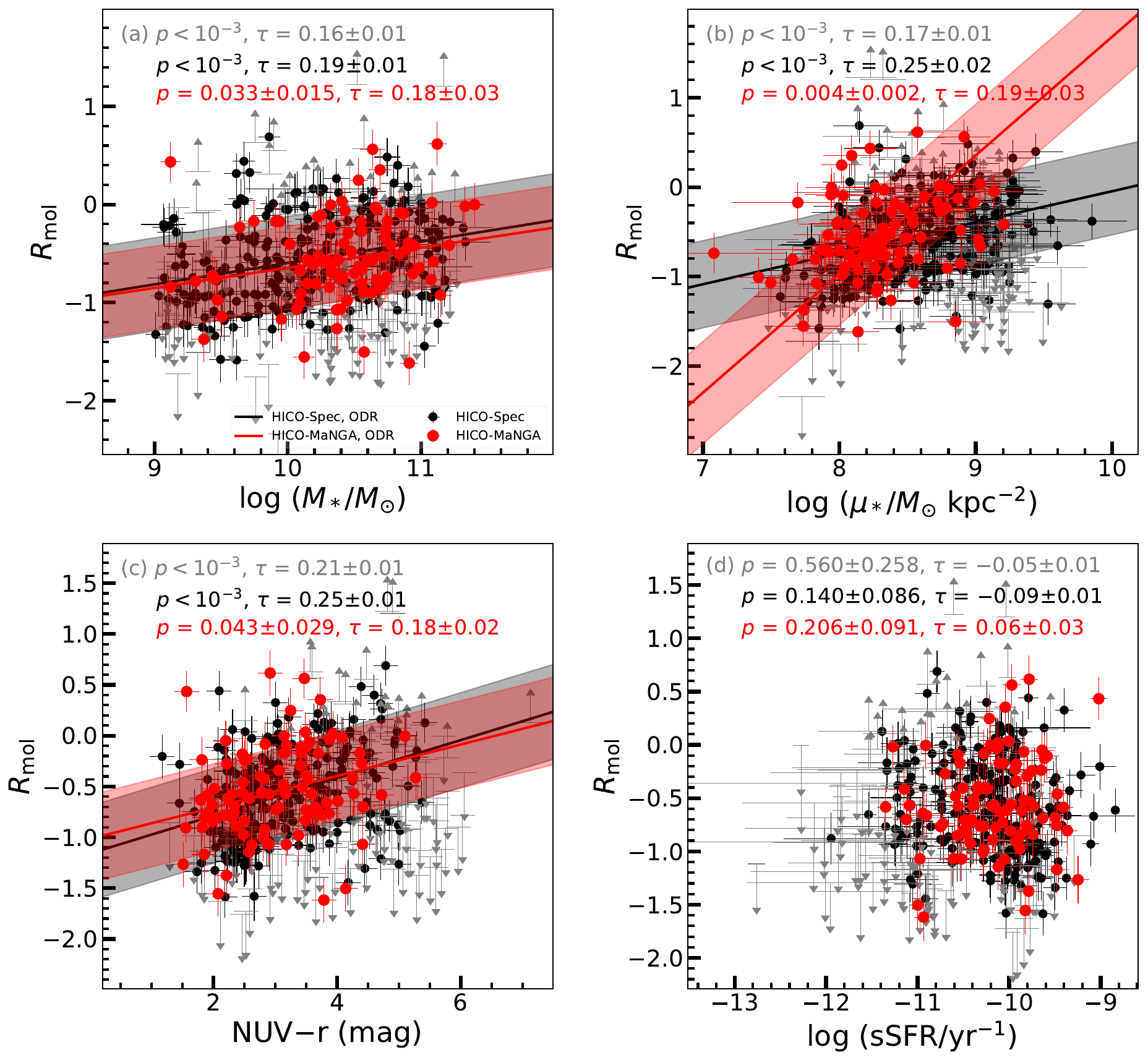}
\caption{The molecular-to-atomic gas ratio as a function of the log \mstar\, (panel a), log $\mu_*$ (panel b), NUV$-$r color (panel c), and sSFR (panel d). In each panel, the black (detection) and grey (non-detection) symbols are galaxies from HICO-Spec, the red symbols are galaxies from the HICO-MaNGA sample. The Kendall's $\tau$ coefficient and $p$ value are shown in the upper left corner (grey: galaxies in HICO-Spec with at least one detection in \HI\ or CO; black: galaxies in HICO-Spec with both detections in \HI\ and CO; red: HICO-MaNGA). The \HI\ or CO non-detection is shown as up and down arrows. The best-fit linear relations based on the orthogonal distance regression are shown as solid lines in panels (a)--(c) with the shaded region denoting the 1$\sigma$ scatter of the fit.}
\label{fig:Rmol}
\end{figure*}

We employ three types of BPT diagrams ([N II], P1P2, and [S II]) to classify the ionization state based on the spatially resolved IFU data from MaNGA. MaNGA~8155-6101 is classified as an AGN host based on BPT diagrams (panels i-n of Figure~\ref{fig:absorption}), and the fraction of AGN or Seyfert pixels decreases from $\sim$100\% to 50\% with the transition from [N II], P1P2, to [S II] BPT diagrams. The high ionization states is likely due to perturbations fuelling an AGN or driving shocks \cite{Kewley2019ARA&A..57..511K}.

The possible overall scenario is that galaxy interaction disturbs the gas distribution in the galaxy outskirt and then results in gas inflow and turbulent motion, which in turn triggers the AGN \cite{Volonteri2003ApJ...582..559V, Goulding2018PASJ...70S..37G, Steffen2023ApJ...942..107S} within MaNGA~8155-6101. To gain further insights into the gas distribution and kinematics under the effects of merger and AGN, future spatially-resolved \HI\ observations may provide valuable details of the gas kinematics in the circum-galactic medium, intergalactic medium, and around the central black hole.

\section{Atomic-to-Molecular Gas Conversion} 
\label{sec:HIH2}

The atomic-to-molecular gas conversion is predominantly facilitated in regions with high mid-plane pressure, weak ultraviolet radiation fields, and high metallicity, which holds both theoretically \cite{Elmegreen1993ApJ...411..170E, Krumholz2009ApJ...693..216K} and observationally \cite{Leroy2008AJ....136.2782L}. Intriguingly, a weak correlation between \MHI\ and $M_{\rm H_2}$ in nearby galaxies suggests that the physical conditions governing the atomic-to-molecular gas conversion are non-trivial \cite{Saintonge2011MNRAS.415...32S, Catinella2018MNRAS.476..875C}. To further constrain the conditions that influence the gas conversion, we carefully scrutinize whether the molecular-to-atomic gas ratio ($R_{\rm mol}\equiv$ log $M_{\rm H_2}$/\MHI) depends on various physical properties. 

\subsection{Stellar Mass Distribution and Star Formation}
\label{subsec:rmol-mustar}
 We investigate the dependence of $R_{\rm mol}$ on \mstar, $\mu_*$, NUV$-$r color, and sSFR in Figure~\ref{fig:Rmol} using the HICO-MaNGA sample in comparison with the HICO-Spec sample (Figure~\ref{fig:Rmol}). Galaxies with at least one detection in \HI\ or CO are shown. The NUV$-$r color is obtained from the NASA-Sloan Atlas (NSA) catalog \cite{Blanton2011AJ....142...31B}, and its typical uncertainty is 0.2 dex \cite{Wyder2007ApJS..173..293W}. The $p$ value\footnote{Following convention, we consider two parameters are correlated if p $<$ 0.05.} and Kendall's $\tau$ coefficient indicate that $R_{\rm mol}$ is correlated with $\mu_*$, less tightly but still correlated with \mstar\, and NUV$-$r color, and not related to the sSFR. The best-fit linear relations in panels (a-c) of Figure~\ref{fig:Rmol} are derived from the orthogonal distance regression. The typical uncertainty of \HI\ and CO detection is 0.1 dex for log \MHI\, and 0.2 dex for log $M_{\rm H_2}$ for our sample. While fitting the linear relations, we assume a typical uncertainty of 0.2 dex for log \MHI\, and 0.4 dex for log $M_{\rm H_2}$ for these non-detections. The uncertainties of $p$-value and Kendall's $\tau$ coefficient are obtained by bootstrap realizations, which considers the uncertainties of both $x$ and $y$ axes.
 
The HICO-Spec sample has a $R_{\rm mol}$ ranges from $-$1.9 to 0.8 with a median value of $\sim -0.7$ \cite{Catinella2018MNRAS.476..875C}. The values of $R_{\rm mol}$ for our sample cover a similar range. 
The HICO-MaNGA sample shows a trend consistent with that of HICO-Spec and xCOLD GASS: $R_{\rm mol}$ increases monotonically with the increasing of \mstar, $\mu_*$, and NUV$-$r color \cite{Saintonge2011MNRAS.415...32S, Bothwell2014MNRAS.445.2599B, Boselli2014A&A...564A..66B, Catinella2018MNRAS.476..875C, Saintonge2022ARA&A..60..319S} with a scatter of $\sim$0.5 dex. 

The positive relation between \mstar\ and $R_{\rm mol}$ is consistent with literature studies.
The typical value of logarithmic molecular-to-atomic gas ratio $R_{\rm mol}$ is 0.6 for S0/Sa galaxies and $-$0.6 for Sd/Sm galaxies \cite{Thronson1989ApJ...344..747T, Young1989ApJ...347L..55Y, Sage1993A&A...272..123S}, which decreases statistically from Sa to irregular galaxies (high mass to low mass, \cite{Young1991ARA&A..29..581Y, Boselli2014A&A...564A..66B}). This weak relation holds true for both constant 
$\alpha_{\rm CO}$ and luminosity-dependent $\alpha_{\rm CO}$ \cite{Bothwell2014MNRAS.445.2599B}. Hydro-dynamical and semi-analytical simulations have also successfully reproduced this relation \cite{Lagos2011MNRAS.418.1649L, Popping2014MNRAS.442.2398P}. Therefore, the efficiency of atomic-to-molecular gas conversion varies along different types of galaxies.

We consider that the dependence of $R_{\rm mol}$ on $\mu_*$ is primarily due to the regulation of mid-plane pressure \cite{Elmegreen1993ApJ...411..170E}, which is within the expectation for a spatially-resolved view: \rmol $\propto  \mu_g^{0.8}(\mu_g+\mu_*v_g/v_*)^{0.8}$ \cite{Krumholz2009ApJ...693..216K}, assuming a thin disk of uniform gas and stars in balance \cite{Blitz2004ApJ...612L..29B}. Here, $\mu_g$ is the gas surface density, and $v_g$ and $v_*$ are the vertical velocity dispersion of the gas and stellar components, respectively.
In a spatially resolved view, the molecular-to-atomic gas surface density ratio is tightly proportional to $\mu_*$ observationally \cite{Leroy2008AJ....136.2782L}. These results support the scenario that
the atomic gas in high-density regions is more easily converted to molecular gas.
This is also
consistent with the observation that massive early-type galaxies have higher values of $R_{\rm mol}$ than late-type galaxies \cite{Boselli2014A&A...564A..66B, Saintonge2022ARA&A..60..319S}.

However, the HICO-MaNGA sample exhibits a steeper slope between $R_{\rm mol}$ and $\mu_*$ compared to that of the HICO-Spec or xCOLD GASS sample. The reason may be that the HICO-MaNGA sample (1) mainly consists of star-forming galaxies, and (2) have log $\mu_*<$ 8.7 \msun\ kpc$^{-2}$. In the bulge-dominated cases with high stellar mass surface densities $\mu_* > 10^{8.7}$ \msun\ kpc$^{-2}$, the cold gas reservoir can be efficiently
converted to stars. Below this $\mu_*$ threshold, higher detection rates are found in both molecular and atomic gas measurements \cite{Catinella2010MNRAS.403..683C, Saintonge2011MNRAS.415...32S}.
As galaxies become more bulge-dominated, both the atomic and molecular gas fractions decrease significantly as well as their detection rates \cite{Saintonge2011MNRAS.415...32S}. The flattening of the relations $\mu_*$ versus $R_{\rm mol}$ for the xCOLD GASS sample is evident above the characteristic thresholds of log\, $\mu_*\sim$ 8.7 \msun\ kpc$^{-2}$, which is mainly due to cold gas depletion in mostly quiescent, early-type galaxies \cite{Saintonge2022ARA&A..60..319S}. The HICO-MaNGA sample is mainly star-forming galaxies and not depleted in cold gas, thus the flattened trend is not observed.

We investigate the relation between \rmol\ and star formation activity in the bottom panels of Figure~\ref{fig:Rmol}, which show a weak and no relation, respectively. NUV$-$r color is an indicator of star formation history or sSFR in galaxies \cite{Wyder2007ApJS..173..293W}. Saintonge \etal\, \cite{Saintonge2011MNRAS.415...32S} also showed \rmol\ is an increasing function of NUV$-$r, but this relation may be a direct result of the strong relation between atomic gas fraction and NUV$-$r \cite{Catinella2010MNRAS.403..683C, Catinella2018MNRAS.476..875C, Zhang2021A&A...648A..25Z}. The NUV$-$r color is sensitive to star formation within several 100 Myrs \cite{Lee2009ApJ...706..599L}, but sSFR is more sensitive to star formation for a longer time scale (several Gyrs), because it quantifies the current star formation activity relative to the existing stellar mass. So the tighter relation between R$_{\rm mol}$ and NUV$-$r color compared with that of sSFR may suggest that the gas conversion is related with the star formation in relatively short timescales.
% So the slightly different trends for NUV$-$r color and sSFR may suggest that the gas conversion is related with recent rather than long timescale star formation.

\begin{figure*}[htbp!]
\centering
\includegraphics[scale=0.5]{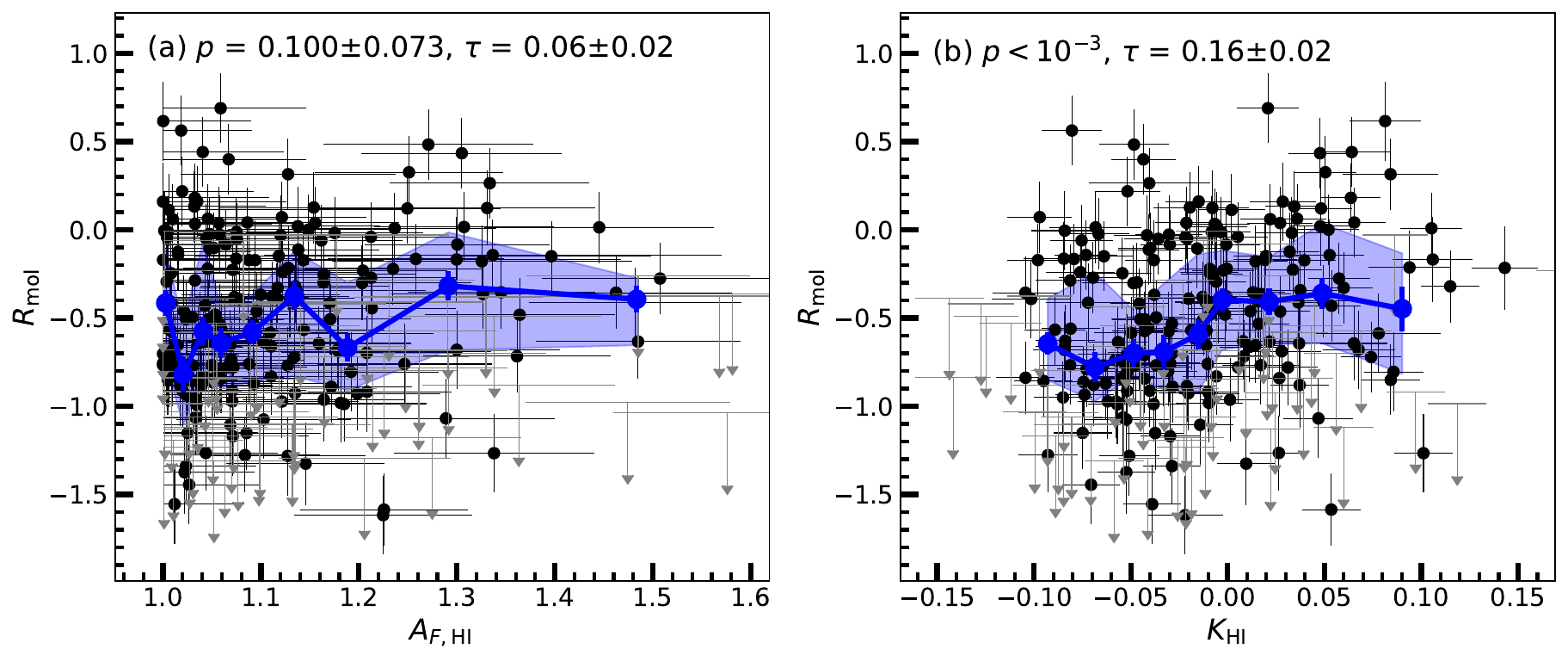}
\caption{The molecular-to-atomic gas ratio as a function of the \HI\ profile asymmetry $A_F$ (panel a) and profile shape $K$ (panel b) for the HICO-Spec sample with \HI\ detection. The black dots are CO detection, and the grey arrows are CO non-detection. The blue circles are the median trend with the standard deviation of median shown as the errorbar, and the blue shaded regions are the 25th and 75th percentiles of the distribution. The $p$ value and Kendall $\tau$ are shown in the upper left corner in each panel.
}
\label{fig:hipara}
\end{figure*}

The HICO-MaNGA and xCOLD GASS samples show no relation between $R_{\rm mol}$ and sSFR (panel d of Figure~\ref{fig:Rmol}). The weak relation is consistent with the results based on 18 detection out of 33 galaxies from the ALMaQUEST sample based on \HI-MaNGA data release 1 \cite{Lin2020ApJ...903..145L}. From late-type to spiral galaxies (sSFR$\gtrsim 10^{-10}$  yr$^{-1}$), the value of $R_{\rm mol}$ increases slowly and then becomes constant in spirals \cite{Boselli2014A&A...564A..66B}.  
The global sSFR is related to the atomic gas fraction (e.g., \cite{Doyle2006MNRAS.372..977D, Huang2012ApJ...756..113H, Zhou2018PASP..130i4101Z}) and molecular gas fraction \cite{Boselli2014A&A...564A..66B} with large scatters. For galaxies on the star-forming main sequence, their variations of $R_{\rm mol}$ are mainly influenced by their \HI\ reservoirs \cite{Catinella2018MNRAS.476..875C}, but the star formation activity relies more on H$_2$ content \cite{Kennicutt2012ARA&A..50..531K}. In addition to the total gas content, the central concentration of cold gas within the optical disk may be more directly related to star formation \cite{Wang2020ApJ...890...63W, Yu2022ApJ...930...85Y}, especially when there is an abundant gas reservoir. Our results suggest that the global conversion from \HI\ to H$_2$ does not significantly influence the global sSFR in non-starburst galaxies.

\subsection{Cold Gas Distribution}
\label{subsec:rmol-hi}
Even though the spatially resolved \HI\ and CO are not simultaneously available for galaxies with a large sample size (e.g., \cite{Leroy2008AJ....136.2782L, Zabel2022ApJ...933...10Z}), the asymmetry and relative concentration of cold gas could be inferred from their global profiles \cite{Yu2020ApJ...898..102Y, Yu2022ApJ...930...85Y}. Meanwhile, spatially resolved gas asymmetry is reflected in asymmetry parameters derived from the global profile \cite{Reynolds2020MNRAS.493.5089R}. The double-horned profile with a deep central trough may suggest centrally depressed \HI\ distribution \cite{Ruffa2023MNRAS.522.6170R}. So the global profile encodes the gas spatial distribution.

We investigate the dependence of $R_{\rm mol}$ on the \HI\ profile asymmetry $A_{\rm F, HI}$ and profile shape $K_{\rm HI}$ for galaxies with \HI\ detection in the xCOLD GASS and HICO-Spec samples (Figure~\ref{fig:hipara}). Our analyses indicate a weak or no relation between $R_{\rm mol}$ and the atomic gas distribution, inferred from either asymmetry $A_{\rm F, HI}$ or profile shape $K_{\rm HI}$. The $p$ value for $R_{\rm mol}$ versus $A_{\rm F, HI}$ is slightly larger than 0.05, thus their relation is not significant. The $p$ value for $R_{\rm mol}$ versus $K_{\rm HI}$ is significantly smaller than 0.05, which suggest that the gas conversion may be related to the \HI\ profile shape $K$.

\begin{figure*}[htbp]
\centering
\includegraphics[scale=0.5]{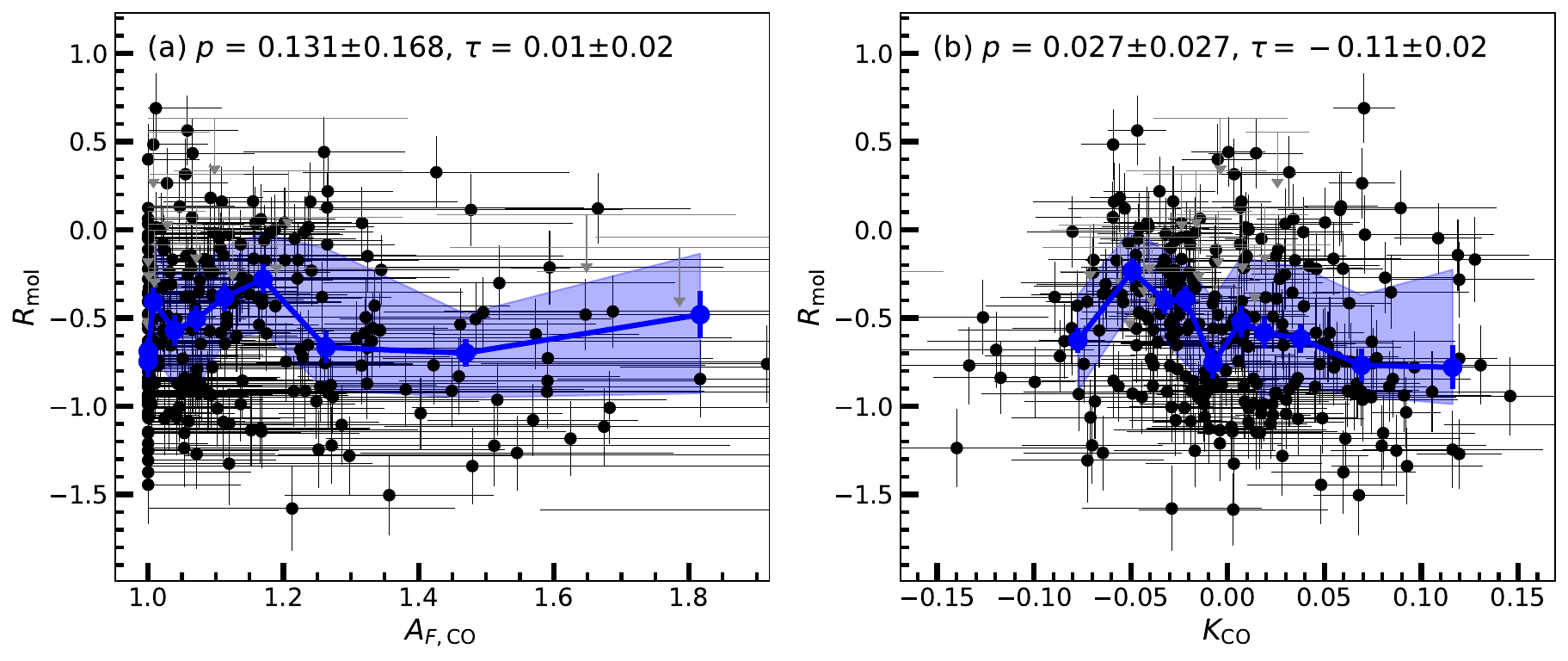}
\caption{The molecular-to-atomic gas ratio as a function of the CO profile asymmetry $A_F$ (panel a) and profile shape $K$ (panel b) for the xCOLD GASS sample with CO detection. The black dots are \HI\ detection, the grey arrows are \HI\ non-detection, and the blue circles shows the median trend. The standard deviation of median is represented by its errorbar, and the blue shaded regions are the 25th and 75th percentiles of the distribution. The $p$ value and Kendall $\tau$ are shown in the upper left corner in each panel.
}
\label{fig:copara}
\end{figure*} 

% The transition of gas from its atomic to molecular form is enhanced
% in regions characterized by high gas surface density \cite{Elmegreen1993ApJ...411..170E}. For example, if the \HI\ surface density is higher than $\sim$ 10 \msun\ pc$^{-2}$ at solar metallicity, molecular hydrogen can persist in a stable state against self-shielding effects \cite{Krumholz2009ApJ...693..216K, Krumholz2009ApJ...699..850K}. 
% Therefore the spatial distribution of cold gas, such as asymmetry and concentration, may be intricately linked to $R_{\rm mol}$. 
Asymmetric features in the distribution and kinematics of \HI\ can be attributed to either internal perturbations, such as stellar \cite{Conselice2000A&A...354L..21C} and active galactic nuclei feedback \cite{Morganti2005A&A...439..521M}, or external perturbations, such as accretion \cite{Sancisi2008A&ARv..15..189S}, stripping \cite{Watts2020MNRAS.499.5205W}, and merging \cite{Bok2019MNRAS.484..582B}. We caution that the profile asymmetry of massive mergers is not higher than that of the control galaxies after correcting for the effects of S/N \cite{Zuo2022ApJ...929...15Z}. Based on high-resolution, spatially-resolved \HI\ distributions in nearby galaxies, Reynolds \etal\, \cite{Reynolds2020MNRAS.493.5089R} argued that gas-removal mechanisms, such as tidal interactions and ram-pressure stripping, are the most probable mechanisms for generating asymmetric \HI\ distributions. The weak relation between $R_{\rm mol}$ and \HI\ asymmetry may suggest that the gas-removal mechanisms do not significantly influence the \HI-to-H$_2$ conversion.
% The efficiency of the conversion from atomic to molecular gas decreases with the increasing of galactic radius \cite{Leroy2008AJ....136.2782L}. Therefore, atomic gas has to be transported to the inner regions of galaxies to facilitate the gas conversion process. 

% Local spiral galaxies have an average \HI\ inflow rate of approximately 0.1 \msun\ yr$^{-1}$ \cite{DiTeodoro2021ApJ...923..220D}, and bars are known to drive the radial inflow of \HI\ at $\sim$ 5 kpc \cite{Eibensteiner2023A&A...675A..37E}. \textbf{The atomic gas inflow is also observed in more than half of the local spiral galaxies} \cite{Schmidt2016MNRAS.457.2642S, DiTeodoro2021ApJ...923..220D}. 
The panel (b) of Figure~\ref{fig:hipara} demonstrates that galaxies with higher $R_{\rm mol}$ tend to exhibit more single-peaked \HI\ profiles in our study, suggesting that \HI\ is relatively more centrally concentrated within the optical disk. Perturbations may drive \HI\ inflow \cite{Schmidt2016MNRAS.457.2642S, DiTeodoro2021ApJ...923..220D, Eibensteiner2023A&A...675A..37E} and promote the atomic to molecular gas conversion \cite{Stark2013ApJ...769...82S}. However, we note that the inclination angle, stellar mass, optical concentration, and gas mass cannot be well controlled in this work as done in Yu, Ho, \& Wang \cite{Yu2022ApJ...930...85Y} to better infer relative gas concentration within the optical disk for nearby galaxies due to the limited sample size.

\HI\ profile shape asymmetry $A_{\rm C, HI}$ shows similar results as that of $A_{\rm F, HI}$, and profile concentration $C_{\rm V, HI}$ also demonstrates comparable trends to $K_{\rm HI}$. Therefore, the \HI\ distribution does not significantly regulate the atomic-to-molecular gas conversion.

Figure~\ref{fig:copara} illustrates the trends of $R_{\rm mol}$ as a function of CO profile asymmetry $A_{\rm F, CO}$ and profile shape $K_{\rm CO}$. The uncertainty of CO profile asymmetry is high, because there are lots of marginal detections with low S/N. Same as Figure~\ref{fig:hipara}, the dependence of $R_{\rm mol}$ on CO distribution is weak or no with $p$=0.131 for $R_{\rm mol}$ versus $A_{\rm F, CO}$ and $p$=0.031 for $R_{\rm mol}$ versus $K_{\rm CO}$. The CO redistribution or asymmetry can be caused by the AGN activity \cite{Stuber2021A&A...653A.172S, Cicone2017A&A...604A..53C} or bar structures \cite{Sormani2019MNRAS.484.1213S}, so the gas conversion does not influence the molecular gas distribution.

Even though the trend is as weak as that of \HI, it is noteworthy that the trends in CO are opposite to those in \HI. The contrasting trends suggest that the distribution and kinematics of \HI\ and CO in galaxies are not identical, which is consistent with previous findings by comparing global profiles of \HI\ and CO  \cite{deBlok2016AJ....152...51D, Roberts2023A&A...675A..78R}. The spatially resolved \HI\ and 
CO asymmetry is weakly correlated for galaxies in the Virgo cluster, but a stronger correlation for the galaxies strongly perturbed by environmental effects \cite{Roberts2023A&A...675A..78R}. Specifically, \HI\ disks tend to be more fragile and extended (such as \cite{Deg2023MNRAS.525.4663D}), but CO is more centrally concentrated within the optical disk. Using global profiles in the xCOLD GASS survey, we find no correlation between the global profile asymmetry and shape of \HI\ and CO, which indicates that  the distribution of atomic and molecular gas in nearby galaxies is not tightly linked. A comprehensive analysis of global \HI\ and CO profiles will be fully investigated in an upcoming paper. 

% The CO distribution seems to be a direct consequence of atomic
% to molecular gas conversion, because .
% Once the molecular gas is formed, it appears to be less perturbed. External perturbations rattle the fragile \HI\ disk and probably drive \HI\ inflow. On the other hand, sometimes the strong radial motion of molecular gas is driven by AGN activity \cite{Cicone2017A&A...604A..53C} or bar structures \cite{Sormani2019MNRAS.484.1213S}, both of which have relatively low incidence rates. Additionally, galaxies with efficient atomic-to-molecular gas conversion \textbf{(higher values of $R_{\rm mol}$)} tend to have more double-horned profiles (lower values of $K$), indicative of their molecular gas tracing the galaxy's rotation and extending within the optical disk. 

\begin{figure*}[ht!]
\centering
\includegraphics[scale=0.4]{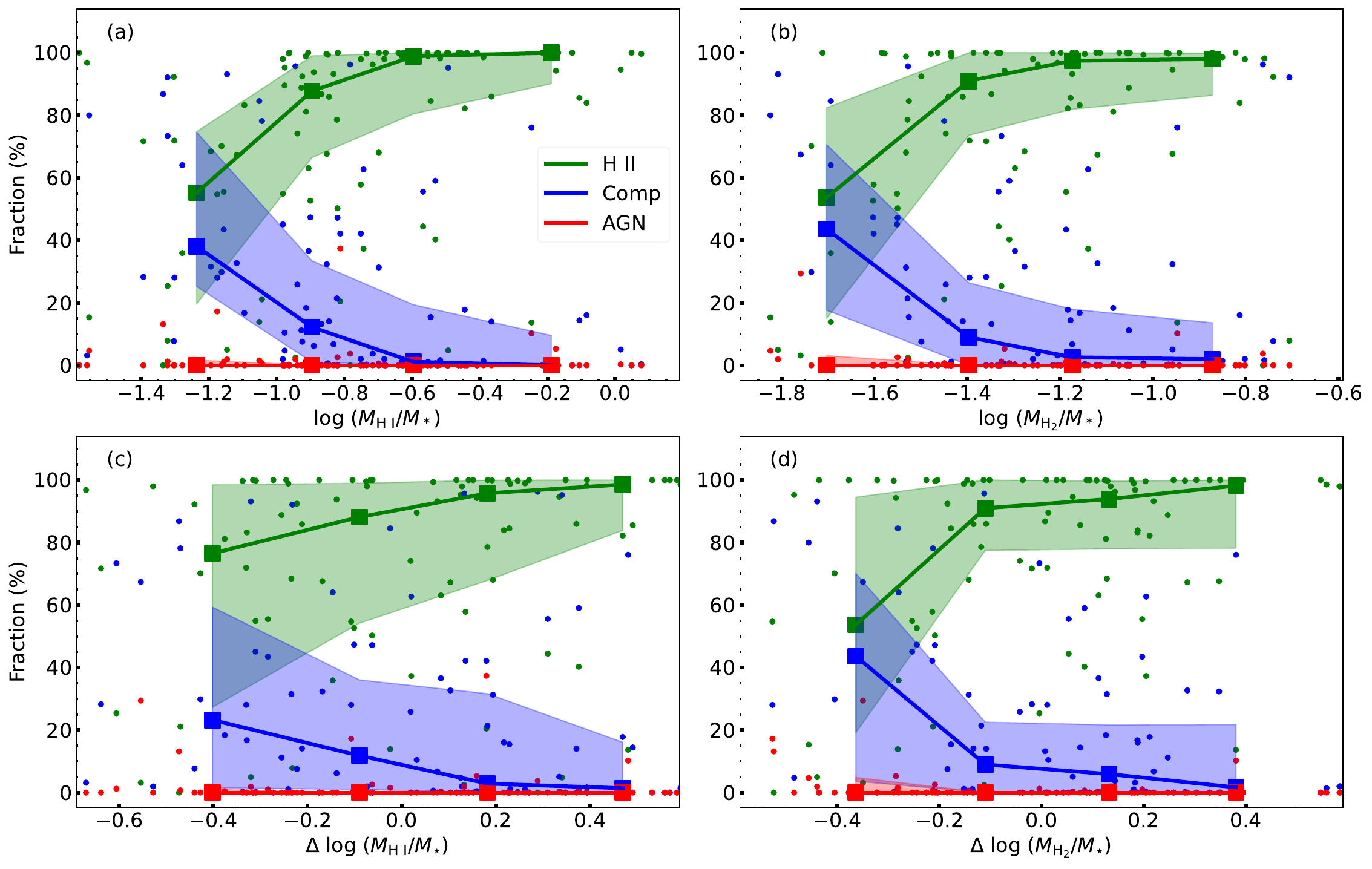}
\caption{The fraction of different optical emission line components classified by P1P2 BPT diagram as a function of the (a) atomic gas fraction, (b) molecular gas fraction, (c) deviation of the atomic gas fraction, and (d) deviation of the molecular gas fraction. In each panel, the small dots show the fraction of each individual galaxy, the squares show the median value of the $x$ and $y$-axis, and the shaded regions represent the 25th and 75th percentiles of the distribution.}
%%%%%%%%%%% std(y)/np.sqrt(N-1), where N is the number of galaxies in each bin
\label{fig:bpt-gas}
\end{figure*}

\subsection{Ionization States} 
\label{subsec:bpt}

We investigate the dependence of the atomic gas fraction log (\MHI/\mstar) and molecular gas fraction log ($M_{\rm H_2}$/\mstar) on the fraction of different ionization states for the HICO-MaNGA sample employing the P1P2 BPT diagram (panel a and b of Figure~\ref{fig:bpt-gas}). 
We have found that the \HII\ region fractions monotonically declines to $\sim$ 50\% with decreasing atomic gas fraction. A similar trend is observed for the molecular gas fraction. As the fraction of atomic or molecular gas increases, the proportion of \HII\ regions in galaxies increases from $\sim$40\% to 100\%. Galaxies with log (\MHI/\mstar)$\geq-$0.8 tend to have \HII\ region fractions of greater than or equal to 80\% (panel a of Figure~\ref{fig:bpt-gas}). Thus the most gas-rich galaxies are predominantly star-forming. 
A similar trend appears in the molecular gas fraction. These trends are consistent across the three types of BPT diagrams: nearly identical results for the [N II] and P1P2 BPT diagrams, whereas the [S II] BPT diagram exhibits a lower fraction of composite regions ($\leq$20\%) and AGN regions ($\leq$5\%). 

To disentangle the dependence of log (\MHI/\mstar) and log ($M_{\rm H_2}$/\mstar) on \mstar, we investigate the relations of the vertical offsets with respect to the median trends as a function of \mstar\ and the fractions of three components (panel c and d of Figure~\ref{fig:bpt-gas}). With the increasing of $\Delta$ log (\MHI/\mstar) and $\Delta$ log ($M_{\rm H_2}$/\mstar), the fraction of emissions from the \HII\ regions increases monotonically, similar to the trends shown in panels (a) and (b). Therefore, the positively-correlated trend indicates that an abundant cold gas reservoir, both atomic and molecular, promotes wide-spread star formation within galaxies. 
% Furthermore, it is evident that the enhancement of SFR is positively related to the \HII\ region fraction. A substantial cold gas reservoir statistically promotes galaxy star formation. 

\begin{figure*}[ht!]
\centering
\includegraphics[scale=0.4]{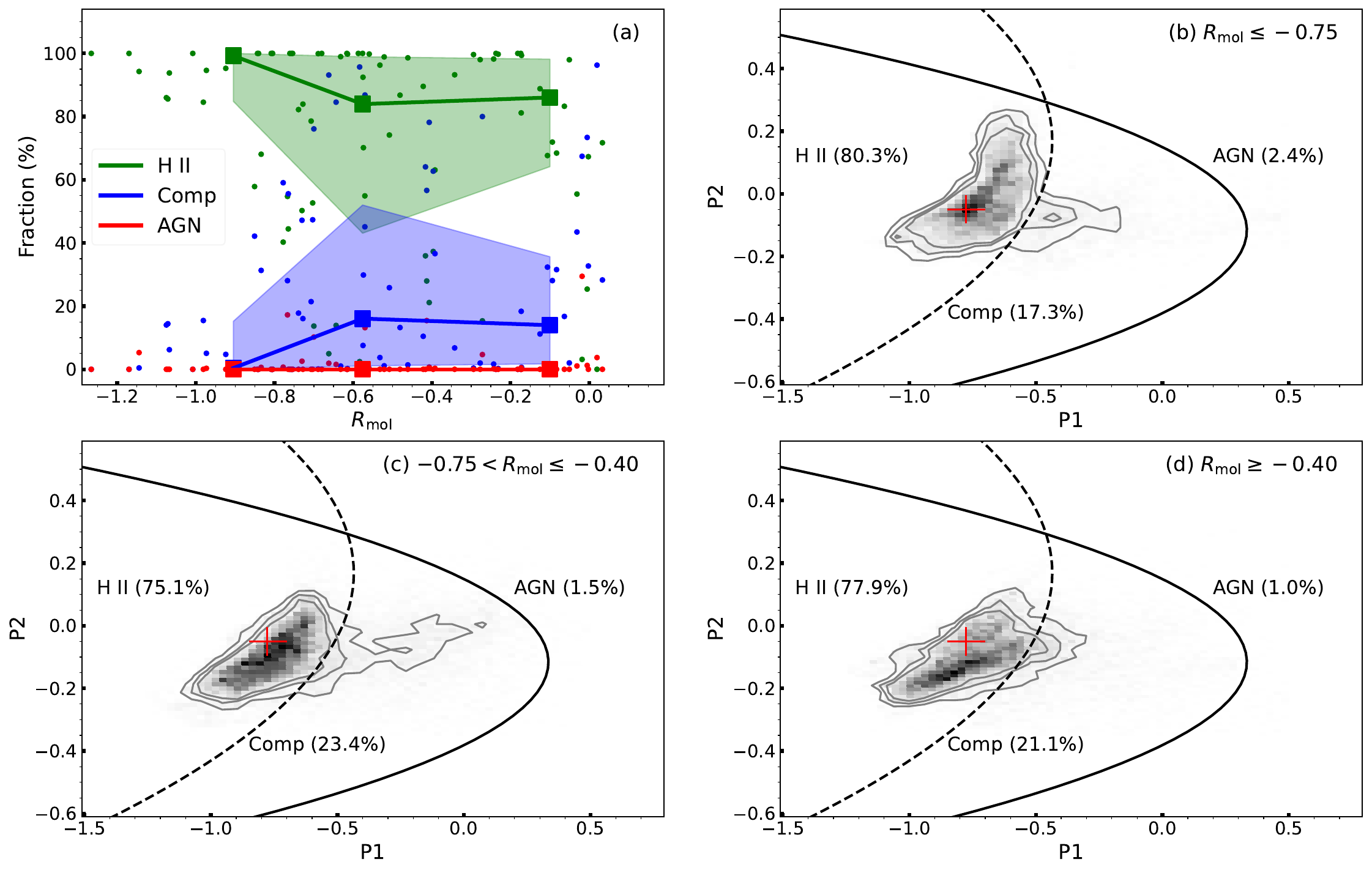}
\caption{Panel (a): The fraction of different optical emission line components classified by P1P2 BPT diagram as a function of the molecular-to-atomic gas ratio ($R_{\rm mol}$). The classification and spatial distribution of three ionization states distinguished by the P1P2 BPT diagram for galaxies with (b) $R_{\rm mol}\leq-$0.75, (c) $-$0.75 $<R_{\rm mol}\leq-$0.40, and (d) $<R_{\rm mol}\geq-$0.40. In panels (b)--(d), the gray scale and contours show the density distribution for all spaxels in each panel, and the red cross shows the density peak of the distribution in panel (b) to highlight the difference of the distribution in three panels. The galaxy numbers in each panels (b)--(d) are the same, and the fractions of \HII\, composite, and AGN-dominated regions are shown.
}
%%%%%%%%%%% std(y)/np.sqrt(N-1), where N is the number of galaxies in each bin
\label{fig:bpt}
\end{figure*}

We investigate how the molecular-to-atomic gas ratio ($R_{\rm mol}$) and the galaxy ionization states influence each other (Figure~\ref{fig:bpt}). All galaxies of HICO-MaNGA have H$\alpha$ equivalent widths larger than 1 $\AA$, both within the field of view and within one effective radius, thus none of them is emission line-less galaxy \cite{Belfiore2016MNRAS.461.3111B}. We adopt the mask of each emission line and require the S/N of each spaxel higher than 10 to minimize the effects of noise \cite{Westfall2019AJ....158..231W}. There are 3 galaxies (MaNGA 8655-12705, MaNGA 12700-12702, and MaNGA 12769-6104) have numbers of spaxel smaller than 100 after adopting the S/N threshold, which may lead to high uncertainties of the fraction of different components, thus we remove them in the following discussion. 
% The inclusion of them does not significantly influence the results.

% that the efficient atomic-to-molecular gas conversion does not significantly boost star formation in galaxies over long timescales, but it is related with the evolution of the galaxy ionization states. 

With the increasing of $R_{\rm mol}$ from $-$1 to 0, the \HII\ region fractions slightly decreases from 100\% to 80\% (panel a of Figure~\ref{fig:bpt}). 
The decreasing of the fraction of the \HII\ regions may be caused by the decreasing of cold gas fraction (see Figure~\ref{fig:bpt-gas}). Because the decreasing trend has a large scatter and the BPT classification is a mixed effect of metallicity and ionization states, we divide the HICO-MaNGA sample into three $R_{\rm mol}$ bins with same galaxy numbers and investigate their distribution in the P1P2 BPT diagram (panels b--d in Figure~\ref{fig:bpt}). The galaxy ionization state increases with the increasing of P1 parameter, and the metallicity decreases with the increasing of P2 parameter \cite{Ji2020MNRAS.499.5749J}. With the increasing of $R_{\rm mol}$, P2 decreases significantly and P1 evolves into two branches. 

From $R_{\rm mol}\leq-$0.75 to $R_{\rm mol}>-$0.75, the fraction of composite regions increases by $\approx$3--6\%. Efficient atomic-to-molecular gas conversion will promote star formation and enhance stellar feedback. The increasing of composite region may be the mixed effects of star formation, shock excitation, and/or AGN activity \cite{Kewley2019ARA&A..57..511K}. On the other hand, the values of P1 decrease significantly with the increasing of $R_{\rm mol}$, but the regions are still classified as \HII-dominated emission. 
The P2 values are anti-correlated with galaxy metallicity, thus higher values of $R_{\rm mol}$ corresponds to higher galaxy metallicity. More researches are vital to reveal detailed physical mechanisms behind the bilateral evolution of P1.

\begin{figure*}[ht!]
\centering
\includegraphics[scale=0.4]{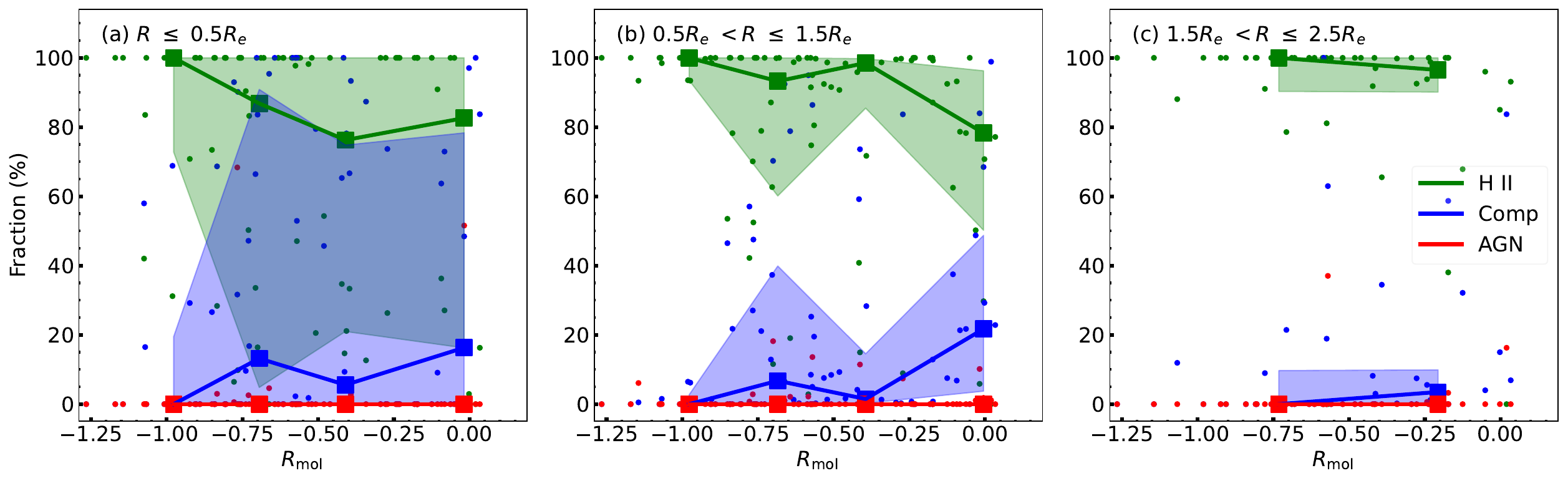}
\caption{The fraction of different components classified by BPT diagram as a function of the molecular-to-atomic gas ratio. From left to right, the galaxy radius range is (a) $R\ \leq$ 0.5$R_e$, (b) 0.5$R_e\ < R\ \leq$ 1.5$R_e$, and (c) 1.5$R_e\ < R\ \leq$ 2.5$R_e$. In each panel, the small dots show the fraction of each individual galaxy, the squares show the median value of the $x$-axis and $y$-axis, and the shaded regions are the 25th and 75th percentiles of the distribution.
}
\label{fig:bpt-radius}
\end{figure*}

The efficiency of atomic-to-molecular gas conversion has been found decreasing monotonically with with the increasing of galactic radius \cite{Leroy2008AJ....136.2782L}, thus we evaluate the dependence of global $R_{\rm mol}$ 
on the fraction of different ionization states within a given galactic radius range in our samples. We divide spatially resolved BPT diagrams (see the example in Figure~\ref{fig:absorption}) into two or three radius ranges: $R\ \leq$ 0.5$R_e$, 0.5$R_e\ < R\ \leq$ 1.5$R_e$, 1.5$R_e\ < R\ \leq$ 2.5$R_e$, where $R_e$ is the effective radius of each galaxy. The results in Figure~\ref{fig:bpt-radius} show that with the increasing of galaxy radius, the fraction of non-star-forming regions decreases from $\sim$ 20\% to 2\%.
It is worth noting that more than 80\% of our sample have an AGN fraction below 5\% in their BPT diagrams, and more than half of our sample have a fraction of the combination of AGN and composite regions lower than 10\%, thus our sample exhibits a low fraction of non-star-forming regions. The non-star-forming emission could be mainly caused by AGN or shock activity \cite{Kewley2019ARA&A..57..511K}. The HICO-MaNGA sample contains 15 galaxies with a weighted AGN and composite fraction higher than 15\%, which tend to be AGN hosts \cite{Wylezalek2018MNRAS.474.1499W}. However, only one galaxy, MaNGA 7977-9101, has a typical H$\alpha$ velocity dispersion higher than 150 km/s, which could be caused by galactic wind shocks or AGN feedback \cite{DAgostino2019MNRAS.485L..38D, Kewley2019ARA&A..57..511K}. Thus, it is difficult to distinguish the main physical drivers for the decrease in the fraction of star-forming regions with the increasing of $R_{\rm mol}$ using the current data.

% It is worth noting that our sample is not specifically selected as AGN host galaxies and exhibits a low fraction of non-star-forming regions, suggesting that AGN feedback may not be the dominant factor contributing to non-star-forming emission in our sample. On the other hand, galaxy star formation and the corresponding stellar feedback decrease with the increasing of galaxy radius. Therefore, Figure~\ref{fig:bpt-radius} may indicate that stellar feedback play a minor role in atomic-to-molecular gas conversion for our sample.

\begin{figure*}[ht!]
\centering
\includegraphics[scale=0.35]{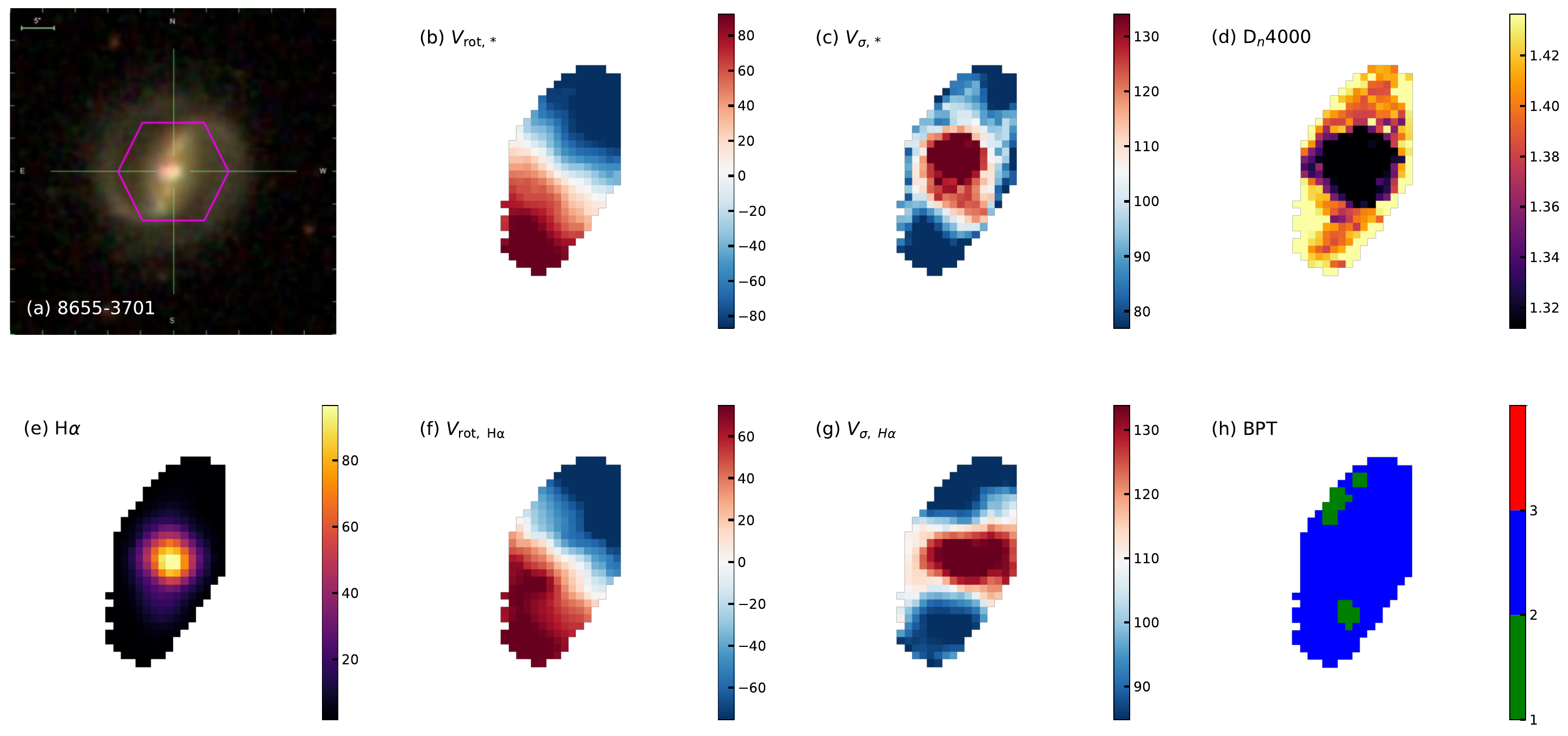}
\caption{The physical properties of MaNGA~8655-3701: panel (a) shows the SDSS composite image, panel (b) and (c)  are the velocity field and velocity dispersion of the stellar component, panel (d) is image of $D_{\rm n}$4000, panels (e)--(g) show the H$\alpha$ image, velocity field, and velocity dispersion, and panel (h) presents the spatially resolved BPT diagram. In panel (h), the red, blue, and green spaxels showing the AGN, composite, and \HII\ regions, respectively.
}
\label{fig:example}
\end{figure*}

% \subsection{Galaxies with high Molecular Gas Ratio but Low \HII\ Contribution}
% \label{subsec:outliers}
We specifically examined 5 galaxies characterized by a high molecular-to-atomic gas ratio and a low fraction of \HII\ regions ($R_{\rm mol}\geq -$ 0.3 and the fraction of \HII\ regions $<$ 50\%): GASS 4030, GASS 11071, MaNGA~8081-12703, MaNGA~9194-3702, and MaNGA~8655-3701. 
These galaxies exhibit distinct characteristics: bright H$\alpha$ core and an old stellar population (high values of $D_n$4000) surrounding the H$\alpha$ core. Their main disks show significant AGN or composite emission. These can be shown in Figure~\ref{fig:example}, where we show the images of the optical, $D_n$4000, H$\alpha$, and BPT of MaNGA~8655-3701: a typical galaxy among these outliers. Both the stellar and H$\alpha$ components display a rotation pattern, and its bright central core has high velocity dispersion. The relative distribution of \HI\ and H$_2$ may be offset, because there is significant difference between their global profiles.

% Meanwhile, the molecular gas distribution shows a \textbf{flat-topped gloabl profile and bright CO core distribution} \cite{Lin2020ApJ...903..145L}, while its \HI\ profile shows a single-Gaussian profile.
% XXXXXXXXXXXXXXXXXXXXX check this again. Hard to conclude.
% The ring structure is associated with composite ionizing gas emission and high values of $D_n$4000 thus older stellar population and composite regions. Despite that MaNGA~7977-9101 has comparable \HI\ and H$_2$ components, its cold gas is primarily distributed in the galaxy outskirt, making it less conducive for directly fueling star formation.

Consequently, we found that a high global value of $R_{\rm mol}$ is not necessarily linked to a high fraction of \HII\ regions. They are separate processes within the baryon cycle in certain cases. The decline in \HII\ region fraction at high molecular-to-atomic ratios can be mainly attributed to high velocity dispersion (see the H$\alpha$ velocity field 
 in Figure~\ref{fig:example}) in the intermediate region (0.5$R_e\ < R\ \leq$ 1.5$R_e$) and extended cold gas distribution. These factors hinder the connection between \HI, H$_2$, and star formation. A similar effect is found that intensive star formation may promote gas velocity dispersion, but nonaxisymmetrical torques can prevent the gas from being gravitationally unstable \cite{Krumholz2018MNRAS.477.2716K}.

\begin{figure*}[ht!]
\centering
\includegraphics[scale=0.45]{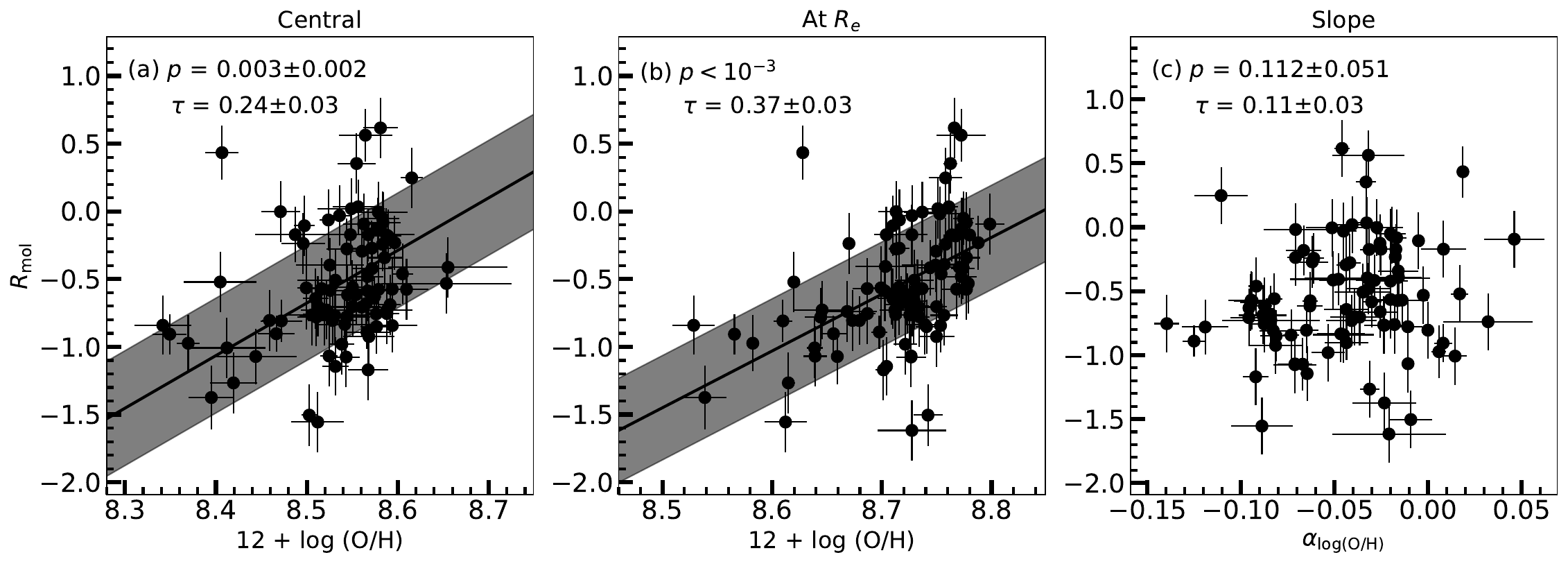}
\caption{The molecular-to-atomic gas ratio as a function of the central oxygen abundance (panel a), the oxygen abundance at one effective radius (panel b), and the slope of oxygen abundance within 0.5 to 2 effective radius (panel c). The oxygen abundance and its slope are derived from O3N2 calibrator taken from Pipe3D \cite{Sanchez2022ApJS..262...36S}. In each panel, the $p$ value and Kendall coefficient $\tau$ as well as their uncertainties are shown in the upper left corner. In panels (a) and (b), the best-fit linear relations based on the Orthogonal distance regression are shown as the black line, with the black shaded region denoting the 1$\sigma$ scatter of the fit.
}
\label{fig:oxygen}
\end{figure*}

\subsection{Metallicity} 
\label{subsec:z}

Metallicity acts as a catalyst in the process of converting  \HI\ to H$_2$ gas, which enhances gas conversion \cite{Elmegreen1993ApJ...411..170E, Krumholz2009ApJ...693..216K}. Gas metallicity serves as a direct indicator of dust \cite{DeVis2019A&A...623A...5D}, and efficient H$_2$ formation occurs in the surface of dust grains \cite{Gould1963ApJ...138..393G}. So metallicity plays a positive role in gas conversion. Boselli \etal\, \cite{Boselli2014A&A...564A..66B} collected optical emission lines and utilized the O3N2 calibration following Pettini \& Pagel \cite{Pettini2004MNRAS.348L..59P}. Their results indicated a correlation between $R_{\rm mol}$ and the oxygen abundance 12+log (O/H) globally. However, the concern about the non-uniformly measured oxygen abundance due to the limitation of the quality of the data could affect the conclusions.

 To address this, we re-investigate the dependence of $R_{\rm mol}$ on the oxygen abundance 12 + log(O/H) by using the data from the MaNGA IFU. The oxygen abundance and its slope are taken from S$\acute{\rm a}$nchez \etal\, \cite{Sanchez2022ApJS..262...36S}, which is determined using the O3N2 $\equiv$ ([O III]/H$\beta$)/([N II]/H$\alpha$) calibration discussed above \cite{Pettini2004MNRAS.348L..59P} for star-forming regions classified by the [N II] BPT diagrams with with EW(H$\alpha) > 3\AA$. The slope of the oxygen abundance is derived from the fitting results obtained between 0.5 and 2.0 $R_e$. As shown in Figure~\ref{fig:oxygen}, the conversion from atomic to molecular gas is enhanced with the increasing of the oxygen abundance at central and one effective radius. 
The oxygen abundance calibrated by the O3N2 index is widely used in high-metallicity regime at solar and super-solar metallicity regime \cite{Pettini2004MNRAS.348L..59P, Marino2013A&A...559A.114M}, where [N II] saturates. The oxygen abundance based on the R23 $\equiv$ ([O II]+[O III]/H$\beta$) and N2 $\equiv$ [N II]/H$\alpha$ calibration \cite{Sanchez2022ApJS..262...36S} returns to similar results. 

So metallicity plays a positive role in gas conversion. Compared to the central metallicity, the metallicity at one effective radius shows a stronger relation with global $R_{\rm mol}$. It suggests that gas conversion predominantly occurs within the optical disk, even though the central region exhibits higher conversion efficiency. We also found that the xCOLD GASS sample represents a positive relation between \rmol\, and global gas-phase metallicity. 

Our data show a weak or no positive relation between $R_{\rm mol}$ and the slope of the oxygen abundance: $p$=0.112 and $\tau$=0.11. 
The positive slopes of oxygen abundance is likely the result of gas accretion \cite{Tissera2022MNRAS.511.1667T}, which dilutes metallicity. 
% Consequently, gas accretion may not directly influence the atomic to molecular gas conversion. 
However, the gradient of oxygen abundance is slightly dependent on stellar mass \cite{Vila-Costas1992MNRAS.259..121V, Tremonti2004ApJ...613..898T, Belfiore2017MNRAS.469..151B}, although EAGLE simulations show no clear trend between them \cite{Tissera2019MNRAS.482.2208T}. The physical drivers of the observed metallicity gradients are complicated, including the radial variations in the SFR \cite{Phillipps1991MNRAS.251...84P, Schonrich2017MNRAS.467.1154S}, gas motions \cite{Queyrel2012A&A...539A..93Q, Pace2021ApJ...908..165P}, and IMF \cite{Guesten1982VA.....26..159G, Martin-Navarro2015ApJ...806L..31M}. Therefore, the lack of correlation between $R_{\rm mol}$ and the oxygen abundance gradient does not necessarily rule out the effects of gas accretion.

The Kendall's $\tau$ coefficient for global $R_{\rm mol}$ versus the oxygen abundance at one effective radius is higher than other relations in Figure~\ref{fig:Rmol}, \ref{fig:hipara}, \ref{fig:copara}, and \ref{fig:oxygen}. Among the dependencies studied in this work, metallicity within the optical disk may be the most important factor regulating the \HI-to-H$_2$ conversion in galaxies, which is consistent with literature studies \cite{Gould1963ApJ...138..393G, Draine2003ApJ...598.1017D}. 

\section{Summary}
\label{sec:sum}

The atomic-to-molecular gas conversion is an important step of the galaxy baryon cycle. We obtained the \HI\ spectra line observations of FAST for the ALMaQUEST sample. We derived 42 emission lines, 1 absorption lines, and 8 non-detection after combining data from Arecibo and GBT. {Our FAST observations are with a noise level of approximately} 0.2 mJy with a velocity resolution of $\sim$ 6 km s$^{-1}$, which is much better than that of \HI-MaNGA. Thanks to the high sensitivity of FAST, the \HI\ detection rate is 80\%, rendering it an ideal instrument for extragalactic H I researches.

Additionally, we compile \HI\ and CO spectra or data from the xCOLD GASS, ALLSMOG, and JINGLE surveys to construct two samples: HICO-Spec and HICO-MaNGA, which contains around 300 and 100 galaxies, respectively. The HICO-MaNGA sample has \HI, CO, and optical IFU observations. This sample is predominantly composed of star-forming galaxies with a stellar mass range of $10^{9.0}-10^{11.5}$\,\msun\ and a redshift range of 0.02 $\lesssim \ z \lesssim$ 0.06.
We analyzed the \HI\ and CO spectra using our recently developed ``curve of growth'' method \cite{Yu2020ApJ...898..102Y, Yu2022ApJS..261...21Y} to measure parameters such as the total flux $F$, profile asymmetry $A_F$, and profile shape $K$. The main results are briefly summarized as below.

\begin{itemize}
    \item[$\bullet$] The molecular-to-atomic gas ratio $R_{\rm mol}$ is positively related to the \mstar, $\mu_*$, and NUV$-$r color, but shows no discernible relation with the sSFR. In contrast to the results of Saintonge \& Catinella \cite{Saintonge2022ARA&A..60..319S}, our sample does not show a flattened trend in the relation of $\mu_*$ and $R_{\rm mol}$. This difference may be attributed to the star-forming nature of our sample and low stellar mass surface density (log $\mu_*<$ 8.7 \msun\ kpc$^{-2}$). Gas conversion depends on mid-plane pressure, which is primarily proportional to surface density, assuming a thin disk with uniform gas and star distribution \cite{Blitz2004ApJ...612L..29B}.
\end{itemize}

\begin{itemize}
    \item[$\bullet$]
The relationship between the $R_{\rm mol}$ and the distribution of \HI\ and CO is weak, while the gas distribution is inferred from the asymmetry and shape of their global profiles. Galaxies exhibiting higher \HI\ asymmetry or more single-peaked \HI\ profiles tend to have higher values of $R_{\rm mol}$, while the opposite trend is observed for CO. Because the relation is weak, we cannot draw definitive conclusions regarding the relationship between the cold gas distribution and \HI-to-H$_2$ conversion.
% These findings suggest that external perturbations may drive the \HI\ inflow, resulting in more single-peaked profiles, and facilitate the atomic-to-molecular conversion, but the distribution of newly formed molecular hydrogen is not significantly disturbed. We caution again that the dependence of $R_{\rm mol}$ on cold gas distribution is weak.
\end{itemize}

\begin{itemize}
    \item[$\bullet$]
With the increasing of $R_{\rm mol}$, the fraction of \HII\ region decreases  within 1.5$R_e$, likely due to the influence of shocks. 
We determine the spatially resolved ionization states using the BPT diagram introduced by Ji \& Yan \cite{Ji2020MNRAS.499.5749J} and examine the dependence of $R_{\rm mol}$ on the fractions of different regions. The fraction of \HII\ regions increases with the increasing of atomic and molecular gas fraction, which is still evident after controlling the effects of stellar mass. Consequently, cold gas reservoirs enhance star formation within galaxies. 
\end{itemize}

\begin{itemize}
    \item[$\bullet$]
However, the fraction of \HII\ regions decreases with the increasing of $R_{\rm mol}$. 
% The decreasing of \HII\ regions fraction is because of the metallicity enhancement and the change in ionization states. 
The decreasing of \HII\ regions fraction is related to the decreasing of cold gas fraction. 
With the increasing of $R_{\rm mol}$, the metallicity increases and the ionization states diverge.
The fraction of composite regions increases by 2\%, which may be due to the stellar feedback. 
% On the other hand, the P1 value decreases significantly within the \HII\ regions, which suggests that the galaxy ionization states decreases due to molecular gas formation.
Therefore, a high molecular-to-atomic gas ratio does not necessarily lead to a high fraction of \HII\ regions, as this is also regulated by galaxy kinematics and gas distribution.
\end{itemize}

\begin{itemize}
    \item[$\bullet$]
Furthermore, galaxies with higher oxygen abundance 12 + log\, (O/H) tend to have higher values of $R_{\rm mol}$, indicating that metellicity acts as a catalyst enhancing atomic-to-molecular gas conversion. The oxygen abundance at $R_e$ proves to be more critical than the central oxygen abundance or oxygen abundance gradients. The slope of oxygen abundance is not tightly related with the global $R_{\rm mol}$, which may suggest that gas accretion does not significantly promote the atomic-to-molecular gas conversion.
\end{itemize}

In summary, the atomic-to-molecular gas conversion mainly depends on stellar mass surface density and metallicity. The most efficient atomic-to-molecular gas conversion occurs in massive or metallicity-rich galaxies. Over a longer timescale, the feedback mechanisms may play a role in the galaxy ionization states, thereby influencing the gas conversion. To gain deeper insights, multi-wavelength observations for stars, ionized gas, atomic gas, molecular gas, and dust with similar resolutions are needed to further constrain the atomic-to-molecular gas conversion in nearby galaxies.

%%%%%%%%%%%%%%%%%%%%%%%%%%%%%%%%%%%%%%%%%%%%%%%%%%%%%%%
%%% Acknowledgements. ??§Ý
%%%%%%%%%%%%%%%%%%%%%%%%%%%%%%%%%%%%%%%%%%%%%%%%%%%%%%%
\Acknowledgements{We sincerely appreciate the constructive suggestions and comments from Lihwai Lin. N.K.Y. thanks the help from Pei Wang. This work was supported by National Science Foundation of China No. 11988101, 11973051, 12041302, 12373012 and U1931110, the China Postdoctoral Science Foundation (2022M723175, GZB20230766), the International Partnership Program of Chinese Academy of Sciences, Program No. 114A11KYSB20210010, National Key\&D Program of China No. 2023YFE0110500, 2023YFA1608004, and 2023YFC2206403, the National Natural Science Foundation of China No. 11903003, and the Ministry of Science and Technology of China No. 2022YFA1605300. This work is also supported by the Young Researcher Grant of Institutional Center for Shared Technologies and Facilities of National Astronomical Observatories, Chinese Academy of Sciences. 
TX acknowledges the support by NSFC No. 11973030. Z.N.L. acknowledges the fellowship of China National Postdoctoral Program for Innovation Talents (grant BX20220301). Hongying Chen is supported by the project funded by China Postdoctoral Science Foundation No. 2021M703236. Di Li is a New Cornerstone Investigator. This work made use of the data from FAST (Five-hundred-meter Aperture Spherical radio Telescope). FAST is a Chinese national mega-science facility, operated by National Astronomical Observatories, Chinese Academy of Sciences. 
This research made use of the NASA/IPAC Extragalactic Database (\url{http://ned.ipac.caltech.edu}), which is funded by the National Aeronautics and Space Administration and operated by the California Institute of Technology. We used Astropy, a community-developed core Python package for astronomy \cite{Astropy2013A&A...558A..33A, Astropy2018AJ....156..123A}.}

%%%%%%%%%%%%%%%%%%%%%%%%%%%%%%%%%%%%%%%%%%%%%%%%%%%%%%%
%%% Conflict of interest. ????????????
%%%%%%%%%%%%%%%%%%%%%%%%%%%%%%%%%%%%%%%%%%%%%%%%%%%%%%%
\InterestConflict{The authors declare that they have no conflict of interest.}

%%%%%%%%%%%%%%%%%%%%%%%%%%%%%%%%%%%%%%%%%%%%%%%%%%%%%%%
%%% Supplements. ????????, ????
%%%%%%%%%%%%%%%%%%%%%%%%%%%%%%%%%%%%%%%%%%%%%%%%%%%%%%%
%\Supplements{}

%%%%%%%%%%%%%%%%%%%%%%%%%%%%%%%%%%%%%%%%%%%%%%%%%%%%%%%
%%% Reference section. ?¦Ï?????
%%% citation in the content using "some words~\cite{1,2}".
%%% ~ is needed to make the reference number is on the same line with the word before it.
%%%%%%%%%%%%%%%%%%%%%%%%%%%%%%%%%%%%%%%%%%%%%%%%%%%%%%%

\end{multicols}
\end{document}